\documentclass[twocolumn]{aastex701}
\usepackage{graphicx}
\usepackage{epsfig}
\usepackage{natbib}
\usepackage{multirow}
\usepackage{amsmath}
\usepackage{textcomp}
\usepackage{graphicx}
\usepackage{comment}
\usepackage{natbib}

\newcommand{\g}{{\rm\thinspace g}}

\newcommand{\s}{{\rm\thinspace s}}

\newcommand{\G}{{\rm G}}
\newcommand{\K}{{\rm K}}

\newcommand{\mgii}{Mg\thinspace{\sc ii}}

\newcommand{\sitwo}{Si\thinspace{\sc ii}}
\newcommand{\sithree}{Si\thinspace{\sc iii}}
\newcommand{\sifour}{Si\thinspace{\sc iv}}

\defcitealias{huberty2024}{H24}
\begin{document}
\linenumbers

\title{CLASSY. XV. Kinematics and Spatial Distributions of Outflows in Local Highly Star-Forming Galaxies}

\correspondingauthor{Mason S. Huberty}
\email{huber458@umn.edu}
\author[0009-0002-9932-4461]{Mason S. Huberty}
\affiliation{Minnesota Institute for Astrophysics, University of Minnesota, 116 Church Street SE, Minneapolis, MN 55455, USA}
\email{huber458@umn.edu}

\author[0000-0003-4166-2855]{Cody A. Carr}
\affiliation{Department of Astronomy, The University of Michigan, 1085 S. University Avenue, West Hall 323 
Ann Arbor, MI 48109, USA}
\affiliation{Center for Cosmology and Computational Astrophysics, Institute for Advanced Study in Physics, Zhejiang University, Hangzhou 310058,  China}
\affiliation{Institute of Astronomy, School of Physics, Zhejiang University, Hangzhou 310058,  China}
\email{CodyCarr24@gmail.com}

\author[0000-0002-9136-8876]{Claudia Scarlata}
\affiliation{Minnesota Institute for Astrophysics, University of Minnesota, 116 Church Street SE, Minneapolis, MN 55455, USA}
\email{mscarlat@umn.edu}

\author[0000-0002-6586-4446]{Alaina Henry}
\affiliation{Space Telescope Science Institute, 3700 San Martin Drive, Baltimore, MD 21218, USA}
\affiliation{Center for Astrophysical Sciences, Department of Physics \& Astronomy, Johns Hopkins University, Baltimore, MD 21218, USA}
\email{ahenry@stsci.edu}

\author[0000-0001-8587-218X]{Matthew Hayes}
\affiliation{Stockholm University, Department of Astronomy and Oskar Klein Centre for Cosmoparticle Physics, AlbaNova University Centre, SE-10691, Stockholm, Sweden}
\email{matthew.hayes@astro.su.se}

\author[0000-0002-9217-7051]{Xinfeng Xu}
\affiliation{Department of Physics and Astronomy, Northwestern University, 2145 Sheridan Road, Evanston, IL, 60208, USA.
}
\affiliation{Center for Interdisciplinary Exploration and Research in Astrophysics (CIERA), Northwestern University, 1800 Sherman Avenue, Evanston, IL, 60201, USA.
}
\email{xinfeng.xu@northwestern.edu}

\author[0000-0003-1127-7497]{Timothy Heckman}
\affiliation{Center for Astrophysical Sciences, Department of Physics \& Astronomy, Johns Hopkins University, Baltimore, MD 21218, USA}
\email{theckma1@jhu.edu}

\author[0000-0002-2644-3518]{Karla Z. Arellano-C\'{o}rdova}
\affiliation{Institute for Astronomy, University of Edinburgh, Royal Observatory, Edinburgh, EH9 3HJ, UK}
\email{K.Arellano@ed.ac.uk}

\author[0000-0002-4153-053X]{Danielle A. Berg}
\affiliation{Department of Astronomy, The University of Texas at Austin, 2515 Speedway, Stop C1400, Austin, TX 78712, USA}
\email{daberg@austin.utexas.edu}

\author[0000-0002-3959-6572]{R. Michael Jennings}
\affiliation{Center for Astrophysical Sciences, Department of Physics \& Astronomy, Johns Hopkins University, Baltimore, MD 21218, USA}
\email{rjenni11@jhu.edu}

\author[0000-0001-9189-7818]{Crystal L. Martin}
\affiliation{Department of Physics, University of California, Santa Barbara, Santa Barbara, CA 93106, USA}
\email{cmartin@physics.ucsb.edu}

\author[0000-0002-8809-4608]{Kaelee S. Parker}
\affiliation{Department of Astronomy, University of Texas at Austin, 2515 Speedway, Austin, TX 78712, USA}
\affiliation{Cosmic Frontier Center, The University of Texas at Austin, Austin, TX 78712, USA}
\email{kaelee.parker@utexas.edu}

\author[0000-0003-3458-2275]{St\'{e}phane Charlot}
\affiliation{Sorbonne Universit\'{e}, CNRS, UMR7095, Institut d'Astrophysique de Paris, F-75014, Paris, France}
\email{charlot@iap.fr}

\author[0000-0002-0302-2577]{John Chisholm}
\affiliation{Department of Astronomy, The University of Texas at Austin, 2515 Speedway, Stop C1400, Austin, TX 78712, USA}
\email{chisholm@austin.utexas.edu}

\author[0000-0002-5659-4974]{Simon Gazagnes}
\affiliation{Department of Astronomy, The University of Texas at Austin, 2515 Speedway, Stop C1400, Austin, TX 78712, USA}
\email{sgazagnes@gmail.com}

\author[0000-0003-3424-3230]{Weida Hu}
\affiliation{Department of Physics and Astronomy, Texas A\&M University, College Station, TX 77843-4242, USA; George P. and Cynthia Woods Mitchell Institute for Fundamental Physics and Astronomy, Texas A\&M University, College Station, TX 77843-4242, USA}
\email{weidahu@ucsb.edu}

\author[0000-0003-4372-2006]{Bethan L. James}
\affiliation{AURA for ESA, Space Telescope Science Institute, 3700 San Martin Drive, Baltimore, MD 21218, USA}
\email{bjames@stsci.edu}

\author[0000-0003-2685-4488]{Claus Leitherer}
\affiliation{Space Telescope Science Institute, 3700 San Martin Drive, Baltimore, MD 21218, USA}
\email{leitherer@stsci.edu}

\author[0000-0003-2589-762X]{Matilde Mingozzi}
\affiliation{Space Telescope Science Institute, 3700 San Martin Drive, Baltimore, MD 21218, USA}
\email{mmingozzi@stsci.edu}

\author[0000-0003-0605-8732]{Evan D. Skillman}
\affiliation{Minnesota Institute for Astrophysics, University of Minnesota, 116 Church Street SE, Minneapolis, MN 55455, USA}
\email{skill001@umn.edu}

\begin{abstract}
Star-forming galaxies drive massive outflows that play an important role in galaxy evolution by regulating feedback and influencing the dynamics of surrounding media. 
Measuring galactic outflow rates is essential for quantifying feedback efficiency and the amount of mass, momentum, and energy deposited into the circumgalactic medium. 
In this paper, we examine 17 galactic outflows from the CLASSY survey with radiative transfer modeling of UV absorption lines presented in \citet{huberty2024}, to study their spatial distributions and kinematic properties.
We study the \sitwo, \sithree, and \sifour\ ionization states that trace the cool and warm phases of the outflows and find that \sitwo\ traces-winds generally behaves differently than the warmer \sithree\ and \sifour\ traced-winds.
We derive the mass, momentum, and energy loading factors, which we find scale inversely proportional to stellar mass.  
We find that our measurements of the mass and momentum loading factors are in agreement with the hydrodynamic FIRE-2 simulations. 
We model the velocity profiles of the winds, with profiles reaching a maximum velocity of $620\rm\ km\ s^{-1}$ on average, in agreement with hydrodynamic simulations from CGOLS.
We also investigate the relationship between outflow properties and the age of the stellar population from SED fitting.  We find that outflows associated with young ($<5\rm\ Myr$) star forming regions are more likely to have a column density dominated by cooler gas and have mass outflow rates which decrease with radius.

\end{abstract}
\keywords{}

\section{Introduction}\label{section1}

Galactic outflows carry matter and energy from the star-forming regions of galaxies into the surrounding circumgalactic medium (CGM) \citep[e.g.][]{shapley2003,martin2005}.
Driven by radiation pressure and mechanical energy deposited by supernovae, stellar winds \citep[e.g.,][]{leitherer1999}, and AGN \citep[e.g.,][]{santoro2020}, they are observed ubiquitously at low and high redshifts
\citep{rupke2018}.
Outflows act to regulate star formation and play an essential role in galaxy evolution \citep[e.g.][]{hopkins2014}. Additionally, these outflows are known to be multi-phase in origin, with temperature regimes often described as cold ($\rm T<10^3$K) \citep{gonzalez2012, falstad2015, herrera2020}, cool ($10^3 < \rm T < 10^5$K), warm ($10^5 < \rm T < 10^7$K) \citep{stocke2013,peebles2014,werk2014, heckman2017, carr2021}, and hot ($\rm T>10^7$K) \citep{strickland1997, laha2018}.

``Down-the-barrel" observations of absorption spectra, in which a galaxy's stellar continuum is used as the background photon source for metal transitions, provides a pathway to understanding the complexities of outflows in local galaxies. 
Unfortunately, these observations produce difficulties in disentangling the outflow's properties with those of the host galaxy's interstellar medium (ISM) \citep{nelson2019}. 
In order to measure outflow rates and other characteristics of outflows from line properties, models rely on a plethora of assumptions such as the spatial distribution and metallicity of the outflows. Misuse or mischaracterization of these assumptions can significantly affect any derived properties of outflows \citep{chisholm2017}. In addition, metal lines represent only a fraction of a percent of the total outflowing gas by mass, and the transformation between metal properties to those of the hydrogen-dominated total outflow mass require numerical conversions of multiple orders of magnitude, which may lead to enormous levels of uncertainties \citep{nelson2019}.

The ultraviolet wavelength regime hosts an abundance of emission and absorption lines of metals \citep[e.g.,][]{xu2022,xu2025,carrcen2025}. These lines primarily trace the cool and warm phases of the outflows \citep{werk2013,tumlinson2017,chisholm2017}, which are thought to carry a large portion of the outflow mass \citep{kima2020, kimb2020, libryan2020}.
However, in order to constrain the properties of outflows via down-the-barrel techniques, high resolution spectroscopy is required.

The physical properties of outflows have also been studied extensively in simulations, which generate observable predictions that can be compared with observations. However, simulations of different physical scales pose distinct challenges, with limited resolution and missing physics often restricting the fidelity of these predictions \citep[e.g.,][]{nelson2019}.
Therefore, capturing the essential properties of outflows is particularly difficult, and degeneracies further complicate interpretation. For example, feedback prescriptions can exhibit significant degeneracies in their cosmological predictions \citep{mitchell2020}. Nevertheless, estimates of mass loading factors are generally considered to be in tension with observations, particularly in low mass galaxies \citep{mcquinn2019}. Understanding these tensions requires larger and more diverse data sets as well as a deeper understanding of how data interpretation translates to simulation output.  
The post-processing of simulation outputs via radiative transfer modeling to produce mock spectra offers a potential solution, allowing one to define how model interpretations translate directly into simulation results \citep{smith2015,gazagnes2024,carrsmith2025,jennings2025}.

In this paper, we use results from semi-analytical radiative transfer modeling of outflows in galaxies from the COS Legacy Archive Spectroscopy Survey \citep[CLASSY;][]{berg2022}, conducted by \citet{huberty2024} (hereafter \citetalias{huberty2024}), to discuss the underlying physics of outflows and to compare their properties with simulations.
CLASSY is a high signal-to-noise and highly resolved data set of down-the-barrel observations of nearby star-forming galaxies in the far-ultraviolet wavelength regime ($\sim 1200$\AA\ - $2000$\AA). These observations allow for extensive modeling of the properties of galactic winds, allowing for measurements of outflow rates, the spatial distribution of outflows, and the kinematics of the gases traced by metal lines. CLASSY sources span a wide range of galactic properties, providing a unique sample of outflows that enables comparison with more uniform outflow-hosting galaxy samples, such as the Low-z Lyman Continuum Survey \citep[LzLCS,][]{flury2022, carrnature2025}, and together these two surveys constitute one of the largest samples of galactic winds.
\citetalias{huberty2024} showed that derived outflow properties depend strongly on the modeling technique used when comparing results inferred from semi-analytical models \citep{carr2023} to empirical methods using the partial covering model \citep[PCM,][]{xu2022}. 
Given this tension, it is worth exploring how the measured properties of outflows stack up against simulations and other surveys of outflows. Semi-analytical radiative transfer modeling of outflows provides a different perspective than the PCM, and provides new constraints on the compositions of outflows that were not available with the PCM. 

This paper is structured as follows: 
in Section~\ref{sec:data+model}, we review the CLASSY data and our modeling procedure.  In Section~\ref{sec:paramaterdiscussion}, we analyze the composition of the outflows. We discuss derived outflow rates and compare them to simulations in Section~\ref{sec:scalingrelations}.  
Our conclusions are outlined in Section~\ref{sec:conclusion}.
This paper assumes a cosmology of $H_0=70$ km s$^{-1}$ Mpc$^{-1}, \Omega_m=0.3,$ and  $\Omega_\Lambda=0.7$.

\section{Data and Modeling}\label{sec:data+model}

The data and modeling procedure are discussed in detail in \citetalias{huberty2024}. For convenience, we summarize the CLASSY sample and the relevant information regarding the computation of the mass, energy, and momentum outflow rates here.

CLASSY is a high-resolution ($R\sim15,000$), far-ultraviolet (FUV) Hubble Space Telescope (HST) legacy survey of 45 star-forming galaxies from a redshift range of $0.002 < z < 0.182$. 
These galaxies were observed with the Cosmic Origins Spectrograph \citep[COS,][]{green2012} covering the FUV between $\sim$1,000\AA\ and $\sim$2,000\AA. 
A detailed description of the data reduction is discussed in \citet{james2022}. 
\citetalias{huberty2024} used the Semi-Analytical Line Transfer Model \citep[SALT, ][]{carr2023} to model the \sitwo\ multiplet (1190\AA, 1193\AA, 1260\AA, 1304\AA, 1527\AA), the \sithree\ singlet (1206\AA), and the \sifour\ doublet (1394\AA, 1403\AA) in 17 CLASSY galaxies. The subset of 17 CLASSY galaxies were chosen as the CLASSY galaxies that had spectral coverage and minimal contamination for all three aforementioned silicon transitions. All relevant \sitwo\ transitions were modeled simultaneously, as were the \sifour\ transitions.

SALT models outflows by solving the radiation transport equation assuming the Sobolev approximation \citep{sobolev,lamers1999} to model the resonant absorption and re-emission (resonant+fluorescent) of photons by the outflows, accounting for emission infilling \citep{prochaska2011, scarlata2015, zhu2015, mauerhofer2021}, omitting thermal and turbulent broadening in the process. The validity of the assumptions underlying SALT modeling has been tested extensively using a second party numerical radiation transport code (RASCAS) in \citet{carr2023} and its predictions verified against simulations in \citet{carrsmith2025}. 

\citetalias{huberty2024} also fit a static ISM component for each silicon absorption line with a gaussian centered at each resonant transition. It is also worth mentioning here that expanding superbubbles have been detected in a subset of CLASSY galaxies, including 6 sources in this subsample \citep{peng2025}. Superbubbles contribute to the broad component (FWHM $\sim200 \rm km s^{-1}$) of hydrogen and metal emission lines, as opposed to the outflow dominated very broad component (FWHM $\sim1200 \rm km s^{-1}$) \citep{martin2024,peng2025}. It is therefore likely that superbubbles contribute to the silicon absorption lines modeled by SALT. However, the low velocity of these bubbles realtive to the strong outflows in this subset of CLASSY galaxies means that a significant portion of absorption due to superbubbles is contained within the static ISM component of SALT, even if the covering fraction of such supperbubbles are large. 
Degeneracies and limitations on SALT are discussed thoroughly in \citet{carr2018,carr2023,carrsmith2025} and in \citetalias{huberty2024}.

Outflow modeling with SALT provides constraints on the following parameters: the opening angle of the outflow ($\alpha$), its orientation angle with respect to the observer ($\psi$, where $\psi=0$ means the outflow is pointed directly towards the observer), the wind velocity and density fields of arbitrary power law (indices $\gamma$ and $\delta$, respectively), the optical depth of the gas ($\tau_0$), the porosity of the outflowing gas ($f_c$, where $f_c=1$ indicated a completely porous outflow), and the launch and terminal velocities of the outflow ($v_0$ and $v_w$) measured at radii $r=R_{SF}$ and $r=R_w$ respectively. 
$R_{SF}$ is taken as the NUV half-light radius for the CLASSY sample \citep{xu2022}.

For each galaxy, the outflow rate of hydrogen is measured assuming that silicon is only in the three observed ionization states, consistent with simulations \citep{carr2025}.  
The mass outflow rate at a given radius in the outflow can be determined for each individual ion: 
\begin{equation}
\label{eq:Mass}
\dot{M}_{\rm ion} (r)=\Omega f_c m_{Si}n_{0} v_0 R_{SF}^2 \left(\frac{r}{R_{SF}}\right)^{(2+\gamma-\delta)},
\end{equation}
where $\Omega=4\pi (1-cos\alpha)$, $n_0=\frac{m c \tau_0 v_0}{\pi e^2 R_{SF}}$ is the number density of silicon at $r=R_{SF}$ \citepalias[see][]{huberty2024}, and $m_{Si}$ is the mass of silicon, which leads to a hydrogen outflow rate of:
\begin{equation}\label{eq:hydromor}
\dot M_{H}(r)=\dot M_{Si}(r)\frac{1}{Z_{Galaxy} Z_{Si,\odot}},
\end{equation} 
where $Z_{Galaxy}$ is the ratio of metals to hydrogen and has units of the solar metallicity, $Z_{Si,\odot}=7.167\times 10^{-4}$ is the silicon mass fraction for the Sun \citep{solar}, and $\dot{M}_{\rm Si}(r) = \dot{M}_{\rm Si\,II}(r) + \dot{M}_{\rm Si\,III}(r) + \dot{M}_{\rm Si\,IV}(r) $. 
As in \citetalias{huberty2024}, we choose the characteristic mass outflow rate for each galaxy to be defined at a radius $r=R_{SF}$.
The derived mass outflow rates and other relevant quantities are listed in Table~\ref{tab:mor}.

\begin{table*}
\begin{center}
 \caption{Table of various outflow rates and related parameters. SFR and Z are taken from \citet{berg2022}. $\dot M_H$ is derived from Equation~\ref{eq:hydromor}. $\eta_H$ is described in Section~\ref{sec:scalingrelations}. $\dot p_H$, $p_*$, $\eta_p$, $\dot E_H$, $E_*$, and $\eta_E$ are described in Section~\ref{sec:pande}. Upper and lower uncertainties are taken as the 16th and 84th percentiles of the marginalized distributions of each parameter. All outflow rates and loading factors in this table are measured at $r=R_{SF}$.} 
 \label{tab:mor}
 %\resizebox{18cm}{!}{%
 \resizebox{\textwidth}{!}{$%
 \begin{tabular}{cccccccccccccccccccc}
  \hline
  Galaxy & log(SFR)  & log($\dot M_H$)  & log($\eta_H$) & log($\dot p_H$)  & log($p_*$)  & log($\eta_p$) & log($\dot E_H$) & log($E_*$)  & log($\eta_E$)&Z/$Z\odot $\\
  & ($M_{\odot} yr^{-1}$) & ($M_{\odot} yr^{-1}$) & &  (dynes) &  (dynes) &  &  (erg/s) &  (erg/s) & \\
  \hline

J0021+0052&$1.07^{+0.14}_{-0.11} $ & $ 1.24^{+0.69}_{-0.03} $ & $ 0.17^{+1.65}_{-0.84} $ & $ 34.03^{+1.64}_{-0.78} $ & $ 34.37^{+0.14}_{-0.11} $ & $ -0.34^{+1.64}_{-0.87} $ & $ 40.75^{+1.62}_{-0.8} $ & $ 42.55^{+0.14}_{-0.11} $ & $ -0.34^{+1.62}_{-0.88} $ & $ 0.30$\\
J0036-3333&$1.01^{+0.19}_{-0.21} $ & $ -0.29^{+0.63}_{-0.22} $ & $ -1.3^{+0.68}_{-0.38} $ & $ 31.64^{+0.55}_{-0.26} $ & $ 34.31^{+0.19}_{-0.21} $ & $ -2.67^{+0.55}_{-0.39} $ & $ 37.72^{+0.48}_{-0.32} $ & $ 42.49^{+0.19}_{-0.21} $ & $ -2.67^{+0.49}_{-0.46} $ & $ 0.33$\\
J0405-3648&$-1.81^{+0.31}_{-0.27} $ & $ 0.57^{+1.95}_{-0.23} $ & $ 2.38^{+1.99}_{-0.54} $ & $ 32.87^{+2.06}_{-0.37} $ & $ 31.49^{+0.31}_{-0.27} $ & $ 1.37^{+2.06}_{-0.58} $ & $ 39.07^{+2.18}_{-0.41} $ & $ 39.67^{+0.31}_{-0.27} $ & $ 1.37^{+2.18}_{-0.63} $ & $ 0.02$\\
J0926+4427&$1.03^{+0.13}_{-0.13} $ & $ 1.54^{+0.42}_{-0.2} $ & $ 0.51^{+0.47}_{-0.31} $ & $ 34.2^{+0.46}_{-0.28} $ & $ 34.33^{+0.13}_{-0.13} $ & $ -0.13^{+0.47}_{-0.34} $ & $ 40.8^{+0.49}_{-0.33} $ & $ 42.51^{+0.13}_{-0.13} $ & $ -0.13^{+0.5}_{-0.39} $ & $ 0.25$\\
J0934+5514&$-1.52^{+0.09}_{-0.07} $ & $ 0.42^{+1.27}_{-0.08} $ & $ 1.94^{+2.09}_{-0.5} $ & $ 32.98^{+2.26}_{-0.55} $ & $ 31.78^{+0.09}_{-0.07} $ & $ 1.2^{+2.26}_{-0.58} $ & $ 39.49^{+2.4}_{-0.63} $ & $ 39.96^{+0.09}_{-0.07} $ & $ 1.2^{+2.4}_{-0.66} $ & $ 0.02$\\
J0938+5428&$1.05^{+0.2}_{-0.17} $ & $ 2.42^{+0.3}_{-0.06} $ & $ 1.37^{+0.94}_{-1.6} $ & $ 35.17^{+0.97}_{-1.15} $ & $ 34.35^{+0.2}_{-0.17} $ & $ 0.82^{+0.97}_{-1.8} $ & $ 41.8^{+1.02}_{-1.19} $ & $ 42.53^{+0.2}_{-0.17} $ & $ 0.82^{+1.02}_{-2.01} $ & $ 0.36$\\
J0940+2935&$-2.01^{+0.42}_{-0.37} $ & $ -0.68^{+1.08}_{-0.14} $ & $ 1.33^{+1.76}_{-0.62} $ & $ 31.48^{+2.2}_{-0.31} $ & $ 31.29^{+0.42}_{-0.37} $ & $ 0.19^{+2.2}_{-0.64} $ & $ 37.58^{+2.63}_{-0.38} $ & $ 39.47^{+0.42}_{-0.37} $ & $ 0.19^{+2.63}_{-0.74} $ & $ 0.09$\\
J1024+0524&$0.21^{+0.14}_{-0.12} $ & $ 1.67^{+0.21}_{-0.01} $ & $ 1.46^{+1.5}_{-1.37} $ & $ 34.33^{+1.46}_{-1.31} $ & $ 33.51^{+0.14}_{-0.12} $ & $ 0.81^{+1.46}_{-1.72} $ & $ 40.89^{+1.42}_{-1.33} $ & $ 41.69^{+0.14}_{-0.12} $ & $ 0.81^{+1.42}_{-1.79} $ & $ 0.14$\\
J1025+3622&$1.04^{+0.14}_{-0.18} $ & $ 1.0^{+0.44}_{-0.12} $ & $ -0.04^{+1.03}_{-0.56} $ & $ 33.25^{+1.16}_{-0.5} $ & $ 34.34^{+0.14}_{-0.18} $ & $ -1.09^{+1.16}_{-0.62} $ & $ 39.5^{+1.25}_{-0.6} $ & $ 42.52^{+0.14}_{-0.18} $ & $ -1.09^{+1.25}_{-0.75} $ & $ 0.28$\\
J1105+4444&$0.69^{+0.28}_{-0.22} $ & $ 1.53^{+0.34}_{-0.04} $ & $ 0.84^{+1.03}_{-0.53} $ & $ 33.67^{+1.13}_{-0.37} $ & $ 33.99^{+0.28}_{-0.22} $ & $ -0.32^{+1.13}_{-0.52} $ & $ 39.73^{+1.26}_{-0.39} $ & $ 42.17^{+0.28}_{-0.22} $ & $ -0.32^{+1.26}_{-0.54} $ & $ 0.35$\\
J1112+5503&$1.6^{+0.2}_{-0.25} $ & $ 2.38^{+0.15}_{-0.09} $ & $ 0.78^{+0.34}_{-0.5} $ & $ 34.7^{+0.33}_{-0.3} $ & $ 34.9^{+0.2}_{-0.25} $ & $ -0.2^{+0.36}_{-0.47} $ & $ 41.02^{+0.37}_{-0.32} $ & $ 43.08^{+0.2}_{-0.25} $ & $ -0.2^{+0.39}_{-0.49} $ & $ 0.58$\\
J1150+1501&$-1.33^{+0.29}_{-0.23} $ & $ -0.72^{+0.59}_{-0.35} $ & $ 0.61^{+1.27}_{-0.56} $ & $ 31.14^{+1.86}_{-0.32} $ & $ 31.97^{+0.29}_{-0.23} $ & $ -0.83^{+1.86}_{-0.48} $ & $ 37.39^{+2.19}_{-0.63} $ & $ 40.15^{+0.29}_{-0.23} $ & $ -0.83^{+2.19}_{-0.89} $ & $ 0.28$\\
J1200+1343&$0.75^{+0.2}_{-0.16} $ & $ 0.67^{+0.1}_{-0.05} $ & $ -0.08^{+0.28}_{-0.35} $ & $ 32.44^{+0.24}_{-0.33} $ & $ 34.05^{+0.2}_{-0.16} $ & $ -1.61^{+0.29}_{-0.41} $ & $ 38.15^{+0.28}_{-0.37} $ & $ 42.23^{+0.2}_{-0.16} $ & $ -1.61^{+0.32}_{-0.46} $ & $ 0.37$\\
J1314+3452&$-0.67^{+0.23}_{-0.55} $ & $ -0.47^{+0.49}_{-0.03} $ & $ 0.2^{+1.21}_{-1.01} $ & $ 31.47^{+1.52}_{-0.38} $ & $ 32.63^{+0.23}_{-0.55} $ & $ -1.16^{+1.52}_{-1.11} $ & $ 37.33^{+1.84}_{-0.43} $ & $ 40.81^{+0.23}_{-0.55} $ & $ -1.16^{+1.84}_{-1.32} $ & $ 0.37$\\
J1359+5726&$0.42^{+0.2}_{-0.14} $ & $ 1.12^{+0.23}_{-0.02} $ & $ 0.7^{+1.3}_{-0.53} $ & $ 32.96^{+1.58}_{-0.46} $ & $ 33.72^{+0.2}_{-0.14} $ & $ -0.76^{+1.58}_{-0.53} $ & $ 38.94^{+1.66}_{-0.73} $ & $ 41.9^{+0.2}_{-0.14} $ & $ -0.76^{+1.66}_{-0.85} $ & $ 0.19$\\
J1428+1653&$1.22^{+0.26}_{-0.19} $ & $ 1.0^{+0.27}_{-0.1} $ & $ -0.22^{+0.43}_{-0.43} $ & $ 33.1^{+0.4}_{-0.49} $ & $ 34.52^{+0.26}_{-0.19} $ & $ -1.42^{+0.44}_{-0.63} $ & $ 39.2^{+0.47}_{-0.58} $ & $ 42.7^{+0.26}_{-0.19} $ & $ -1.42^{+0.49}_{-0.74} $ & $ 0.44$\\
J1429+0643&$1.42^{+0.11}_{-0.17} $ & $ 1.08^{+0.3}_{-0.14} $ & $ -0.34^{+0.39}_{-0.32} $ & $ 33.73^{+0.37}_{-0.25} $ & $ 34.72^{+0.11}_{-0.17} $ & $ -0.99^{+0.37}_{-0.34} $ & $ 40.31^{+0.41}_{-0.3} $ & $ 42.9^{+0.11}_{-0.17} $ & $ -0.99^{+0.41}_{-0.39} $ & $ 0.26$\\

\hline
  
 \end{tabular} $}
 %}
 \end{center}
\end{table*}

\section{Outflow Composition}\label{sec:paramaterdiscussion}

We first explore how the outflow composition changes between each ionization state of silicon.
The three ionization states of silicon that we observe in the CLASSY spectra probe gases at different temperatures, with ionization potentials as follows: 
\sitwo: 8.1-16.3 eV, \sithree: 16.3-33.5 eV, and \sifour: 33.5-45.1 eV.

\subsection{Spatial Distributions and Kinematics}\label{sec:propertycomparison}
\begin{figure*}
    \centering
    \includegraphics[width=1.\linewidth]{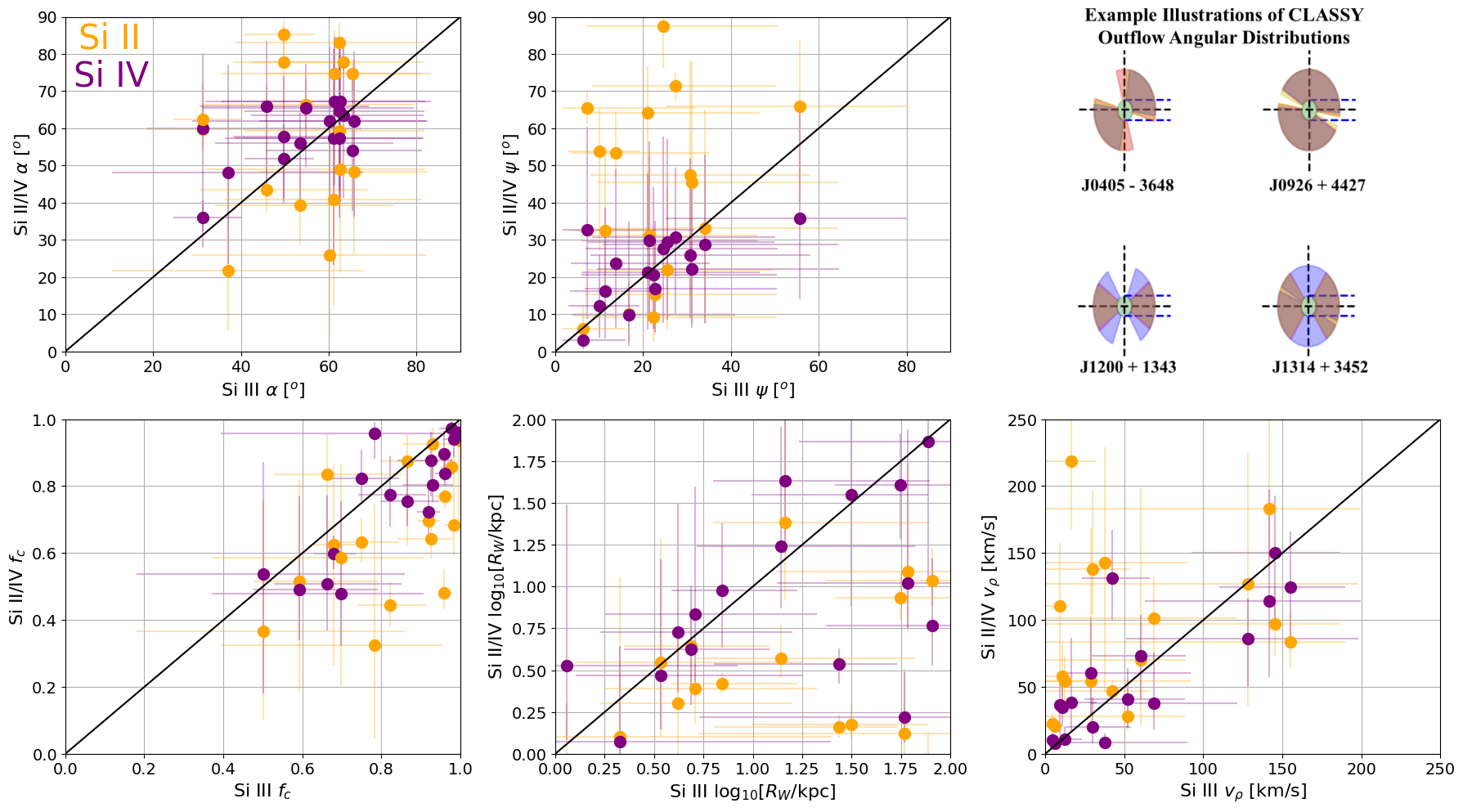}%
    \caption{Comparisons for the opening angle $\alpha$ (upper-left), the orientation angle $\psi$ (upper-center), the porosity $f_c$ (lower-left), the terminal radius $R_W$ (lower-center), and the density weight outflow velocity $v_\rho$ (lower-right) between the different ionization states of silicon. The \sithree\ properties are given on the x-axis, and with the \sitwo\ (\sifour) properties given on the y-axis in orange (purple). The upper right shows 4 example outflow angular distributions oriented with respect to an observer at the right of the page. Brown indicates that all three ionization states overlap at a particular angle; blue, yellow, and red indicate that only \sitwo, \sithree, and \sifour\ respectively are detected at that angle. The cooler \sitwo-traced gas typically has more discrepant spatial and kinematic properties than the warmer \sithree- and \sifour-traced gases.}
    \label{fig:salthistogram}
\end{figure*}

The three different silicon outflow diagnostics indicate that gas at different temperatures exhibits distinct kinematic and physical behaviors.
In Figure~\ref{fig:salthistogram}, we show a comparison of the measured outflow properties between the different ionization states of silicon for the 17 galaxies. In each panel, the measured properties for \sitwo\ (orange) and \sifour\ (purple) on the y-axis are compared with the measured properties for \sithree\ on the x-axis. In the upper-left panel, the opening angle, $\alpha$, is often shown to be large (on average $\alpha\approx58^\circ$ for \sitwo, $\alpha\approx54^\circ$ for \sithree, and $\alpha\approx59^\circ$ for \sifour, where $\alpha=90^\circ$ is a uniform sphere) indicating that few narrow bicones are found in this sample. Additionally, in each galaxy, the measured value of $\alpha$ for \sithree\ and \sifour\ are more closely correlated with each other (with a Pearson $r=0.58$), than between \sithree\ and \sitwo\ ($r=0.17$). This behavior is also seen in the upper-central panel, which shows the comparison for the orientation angle, $\psi$. The Pearson correlation coefficient between \sithree\ and \sifour\ is $r=0.60$ while the correlation coefficient between \sithree\ and \sitwo\ is $r=0.25$. 
An illustration of some example outflow angular distributions are shown in the upper right of Figure~\ref{fig:salthistogram}, where each color represents the outflow in each ionization state (blue for \sitwo, yellow for \sithree, red for \sifour, and brown is where all three states overlap) and $\psi$ is oriented with respect to an observer at the right side of the page. A $\psi=0^\circ$ indicates that the outflows are pointed directly towards the observer (which for example is roughly true for all ionization states of the outflow in J1200+1343). In the top two illustrations, the outflows are measured to have similar $\alpha$ and $\psi$ regardless of ionization state. However, the bottom two examples demonstrate that that the multiple phases of the outflow are not always co-spatial (with significant \sitwo\ deviations from \sithree\ and \sifour) in agreement with the findings in \citet{carr2021}.

In the lower-left panel of Figure~\ref{fig:salthistogram}, we find that the porosity of the outflow, $f_c$, is larger than $0.3$ in all cases. The two higher ionization states, however, tend to have a higher $f_c$ with an average $f_c={0.66}$ for \sitwo, ${0.82}$ for \sithree, and ${0.76}$ for \sifour, indicating that the warmer phases of the outflows are generally more porous than the cooler phases.

The lower central panel shows the comparison for $R_W$, where the radii of these outflows are generally on the order of or slightly larger than those measured in \citet{xu2023} for a similar sample of CLASSY galaxies. The cooler gas traced by \sitwo\ typically extends to smaller distances into the circumgalactic medium than the warmer gas traced by \sithree\ and \sifour. 
It is worth noting here that the preferred low values of $\psi$ for \sithree\ and \sifour\ could be the result of a bias: 
since $R_W$ is typically measured to be larger for the \sithree\ and \sifour, these ions are more likely to suffer emission losses from the COS aperture \citep[e.g.,][]{scarlata2015,carrsmith2025}. 
When the limiting aperture removes emission from the poles of the galaxy, narrow bicones can appear spherical when oriented along the line of sight. The effect of limited aperture on emission is especially evident in the more extended sources, such as J0934+5514, which SALT predicts to have more emission when a larger observing aperture is assumed during the radiative transfer modeling procedure \citep[see also][]{scarlata2015}. 

Finally, in the lower-right panel, we show the comparison for the density-weighted outflow velocity ($v_{\rho}$=$\int^{R_W}_{R_{SF}} v(r)n(r)dr/ \int^{R_W}_{R_{SF}} n(r)dr$, see \citetalias{huberty2024}). 
The typical $v_{\rho}$ of \sitwo\ tends to be larger than its \sithree\ and \sifour\ counterparts. This physically means that for the cool gas, the typical mass per velocity bin is larger at faster velocities than for warmer gas. This is consistent with an idea that mass outflow is dominated by cooler gas. However, this does not mean that all cool gas is faster than all warm gas.

To summarize these results, from both a spatial and kinematic stand point, the gas probed by \sitwo\ behaves differently from the gas probed of \sithree\ and \sifour. The cooler gas typically flows faster, is less porous, and terminates at a smaller radius that its warmer gas counterparts in the CLASSY sample.

\subsection{Makeup of Silicon in the Outflow}\label{sec:makeup}

\begin{figure*}
    \centering
    \includegraphics[width=0.99\linewidth]{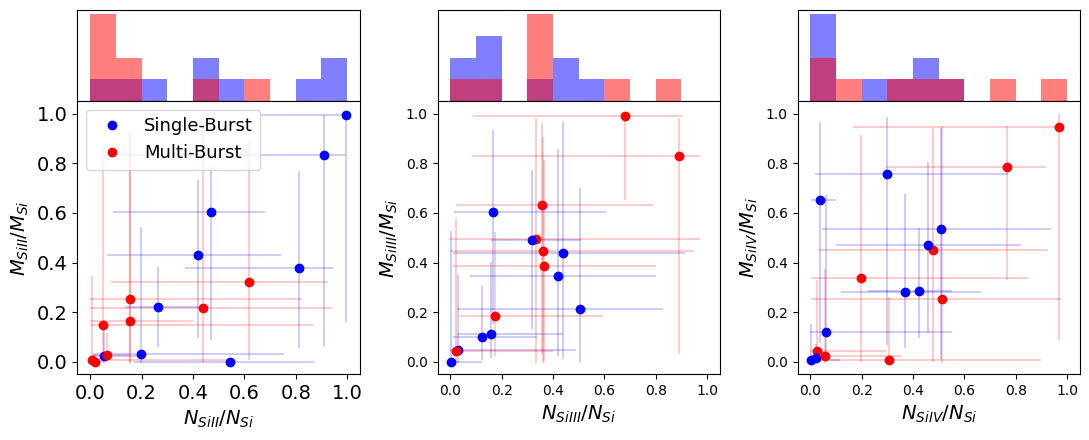}
    \caption{Comparison between the proportion of the outflow that is in the \sitwo\ (left panel), \sithree\ (middle panel), and \sifour\ (right panel) ionization state relative the total amount of silicon in terms of mass (y-axis) and column density (x-axis). Each point is color-coded according to its classification as a younger, single-burst source (blue) or an older, multi-burst source (red). The histograms in the top row reflect the distribution of $N_{Si,ion}/N_{Si}$.}
    \label{fig:ion}
\end{figure*}

We now investigate which ion of silicon dominates the outflows, in terms of column density and total mass.
\citetalias{huberty2024} computed the column density along the line of sight for the three ionization states of silicon. 
The total mass of each silicon ion within the outflow can be found with $M_{ion, total}$=$\int^{R_W}_{R_{SF}} f_c m_{ion} n_0 (\frac{r}{R_{SF}})^{-\delta} 4\pi\ (1-cos(\alpha)) r^2 dr$. 
Figure~\ref{fig:ion} shows the relative contributions of \sitwo\ (left panel), \sithree\ (middle panel), and \sifour\ (right panel) to the total silicon column density (x-axis) compared to each silicon ions contribution to the total silicon outflow mass (y-axis) of silicon for the 17 galaxies. Errors are derived in a Monte Carlo manner, where samples are drawn from the individual \sitwo, \sithree, and \sifour\ column density distributions, and the 16th and 84th percentiles of the resulting $\rm N_{SiII}/N_{Si}$ distribution are taken (and likewise for the mass, $M_{Si}$). 
These measurements indicate that there is no preferred ionization state in the outflows of the CLASSY galaxies, but rather, that some galaxies have outflows which are primarily dominated by \sitwo, while others have outflows in which silicon is predominantly in a higher ionization state.
We find that 5/17 galaxies have no single ionization state making up over 50\% of an outflow's column density (and this number increases to 9/17 if the cutoff is raised to 55\%).

We divide our galaxy sample into two populations: ``single-burst" and ``multi-burst", in the same manner as in \citet{parker2025}. \citet{parker2025} conducted stellar continuum fitting of the CLASSY galaxies, in part to determine the light-weighted ages of the stellar populations enclosed within the COS aperture. This work used estimates of the luminosity-weighted ionizing photon production to divide the CLASSY sample into these two categories: a single-aged population (single-burst), which has a negative correlation between ionizing photon production and age, and a mixed-aged population (multi-burst), which produces large quantities of ionizing radiation over more extended periods. 
In the CLASSY sample, single-burst galaxies tend to have younger ($\lesssim5\rm Myr$) light-weighted populations than the multi-burst ones ($\gtrsim10\rm Myr$), and thus much of the younger, single-burst population may still be in the pre-supernova phase \citep{geen2016}, which may delay efficient ionization of the outflowing gas.
The color of the points and histograms in Figure~\ref{fig:ion} reflect these two subgroups, with blue representing the single-burst sample, and red representing the multi-burst sample.

Galaxies that are dominated by \sitwo\ in terms of both column density and outflow mass tend to be from the younger, single-burst population, whereas galaxies dominated by \sithree\ and \sifour\ tend to come from the older, multi-burst population. To quantify this, we conduct a two sample Kolmogorov-Smirnov (KS) test between the single- and multi-burst sources values of $N_{SiII}/N_{Si}$. We find a KS-statistic of 0.639 with a p-value of 0.047, indicating there is evidence that the underlying single- and multi-burst sources are drawn from different populations.
This emphasizes that all three ionization states of silicon and the gases they trace are physically unique, associated with different underlying stellar populations, and all of this information is required to understand the full picture of outflows.

\section{Outflow Rates}\label{sec:scalingrelations}

\begin{figure*}
    \centering
    \includegraphics[width=1\linewidth]{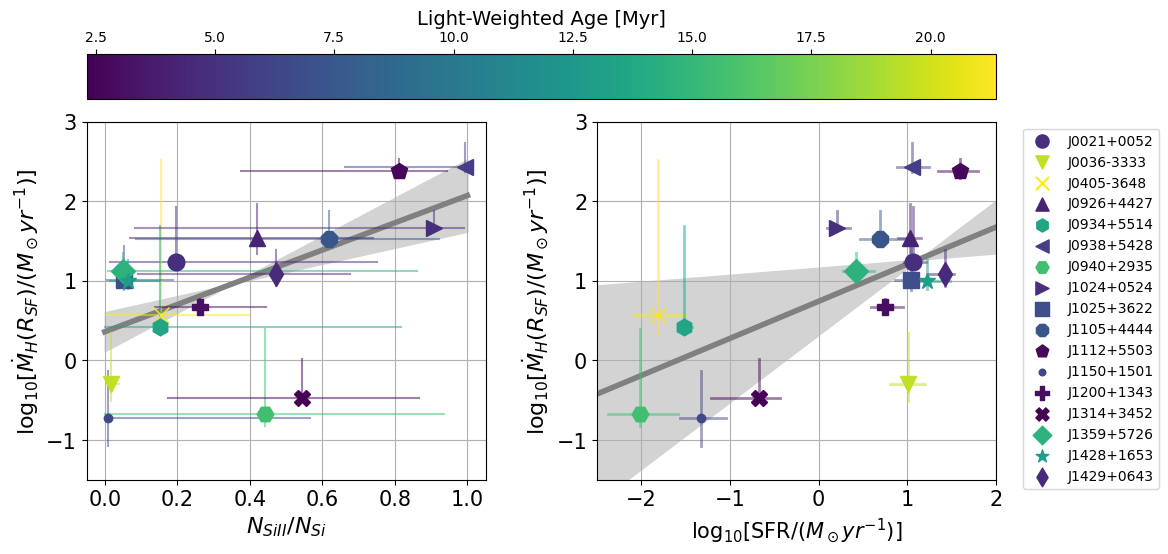}%
    \caption{Left: Comparison between the measured mass outflow rate and the proportion of the outflow column density in \sitwo. The greater proportion of the column density in the cool-gas tracing \sitwo\ ion tends to correlate with the mass outflow rate of hydrogen. Right: Comparison between the measured mass outflow rate and the star-formation rate for the CLASSY galaxies, suggesting a direct correlation between these two galactic properties. In both plots, the galaxies are colored by the light-weighted age of the stellar population within COS, where we find younger galaxies tend to be dominated by cooler gas and have higher mass outflow rates.
    }
    \label{fig:Mdotoutflowrate}
\end{figure*}

Outflow rates allow us to quantify the impact that star-formation driven feedback has on the local environment. 
In the left panel of Figure~\ref{fig:Mdotoutflowrate}, we show the relationship between the mass outflow rate ($\dot M_H$) at $R_{SF}$, and the proportion of the column density of silicon in the outflow that is attributed to \sitwo\ ($N_{SiII}/N_{Si}$). The color of each point corresponds to the light-weighted age of the stellar population contained within the COS aperture \citep[see][]{parker2025}\footnote{The limited COS aperture size may introduce some discrepancy between the stellar population within COS and the galaxy as a whole, particularly for $J0405-3648$, $J0940+2935$, and $J1105+4444$ which have an $R_{SF}>2R_{COS}$. However, for the remainder of the sample, the COS aperture encloses the vast-majority of the star-formation.}.
In outflows that tend to be dominated by \sitwo, $\dot M_H(R_{SF})$ is found to be larger, and this occurs primarily in galaxies with younger light-weighted stellar populations. 
Theoretical models of outflows have shown that mass outflow is conducted primarily through the cool phase of winds \citep[while energy loading is primarily conducted through the hot phase of winds, see][]{kima2020,kimb2020,libryan2020}. Given that \sitwo\ traces the cooler gas, it seems consistent that the mass outflow rate correlates with a higher dominance of \sitwo. 
We may be observing a two-stage burst phenomenon where young galaxies start driving neutral gas (\sitwo) and clear out the ISM.  After this process has evolved, if there are still massive stars, then these stars can then more effectively ionize the surrounding medium \citep[in a one-two punch manner similar to the one outlined in][]{flury2025}. Thus older galaxies would have weaker mass outflow, but with more highly ionized winds.

There are two primary outliers to this trend that are worth commenting on: J1314+3452 and J0940+2935. J1314+3452 is unusual in that while \sitwo\ accounts for much of the silicon in terms of column density, in terms of total outflow mass, \sithree\ and \sifour\ dominate. This is because the terminal radius ($\rm R_W$) of the \sitwo\ outflow is measured to be much smaller than that of the \sithree\ and \sifour\ outflows (the most extreme case of this in the sample). Therefore $\dot M$ is dominated by the higher ionization states at larger radii, while \sitwo\ only dominates for very small radii. J0940+2935 is not an atypical source as such, but it is worth mentioning that it is the source with the largest uncertainty on $N_{SiII}/N_{Si}$.

In the right panel of Figure~\ref{fig:Mdotoutflowrate} we show the relationship between the hydrogen mass outflow rate, $\dot M_{H}$, and the star-formation rate (SFR). These two quantities have been observational found to be correlated in low-redshift galaxies \citep[e.g.][]{heckman2015,xu2022}. This relation is perhaps intuitive as galaxies with more active star formation have a large number of supernovae and stellar winds from large stars - principal drivers of outflows. However, here we also look at the correlation in terms of the light-weighted age: the youngest galaxies tend to have both larger SFRs and $\dot M_{H}$. In these young sources ages (pre-supernova), stellar winds and ionizing radiation from massive stars begin to drive the outflow. When the most massive stars collapse as supernova ($\sim3-8$ Myr), the outflow becomes more highly ionized. 
The oldest galaxies are more likely to have both lower SFR and lower $\dot M_{H}$ compared to the younger population. This is in contrast to what \citet{carrnature2025} found in a select group of galaxies in the Low-z Lyman Continuum Survey \citep[LzLCS,][]{flury2022}. LzLCS is a survey of young star-forming galaxies at $z\sim0.3$ (at a higher-z than CLASSY), of which 6 compact, highly star-forming galaxies were found to have $\dot M_{H}$ that increases with time over similar time-scales to CLASSY. However the galaxies in the \citet{carrnature2025} sample was limited to a narrower range of stellar masses, SFRs, and metallicities, as well as being compact sources. The CLASSY sources are less compact and span a wider range of stellar parameters. The lower compactness of these CLASSY sources may imply that early ionization is less effective, leading to very dense radiation-driven winds at early times \citep[e.g.][]{diamond2012}. Once supernova feedback becomes dominant, the winds transition to a hot, thermalized phase.
At the largest ages in the CLASSY sample ($>15$Myr) the most-massive stars have already gone off as supernova, and the shocked gas may be in the very hot phase $T>10^7\rm K$, which is hotter than the gas traced by \sifour. Therefore the mass outflow rate may be larger than measured from the UV silicon lines for these older sources. What is clear is that age, star-formation, and mass all have a role to play in outflow dynamics.

\begin{figure*}
    \centering
    \includegraphics[width=0.9\linewidth]{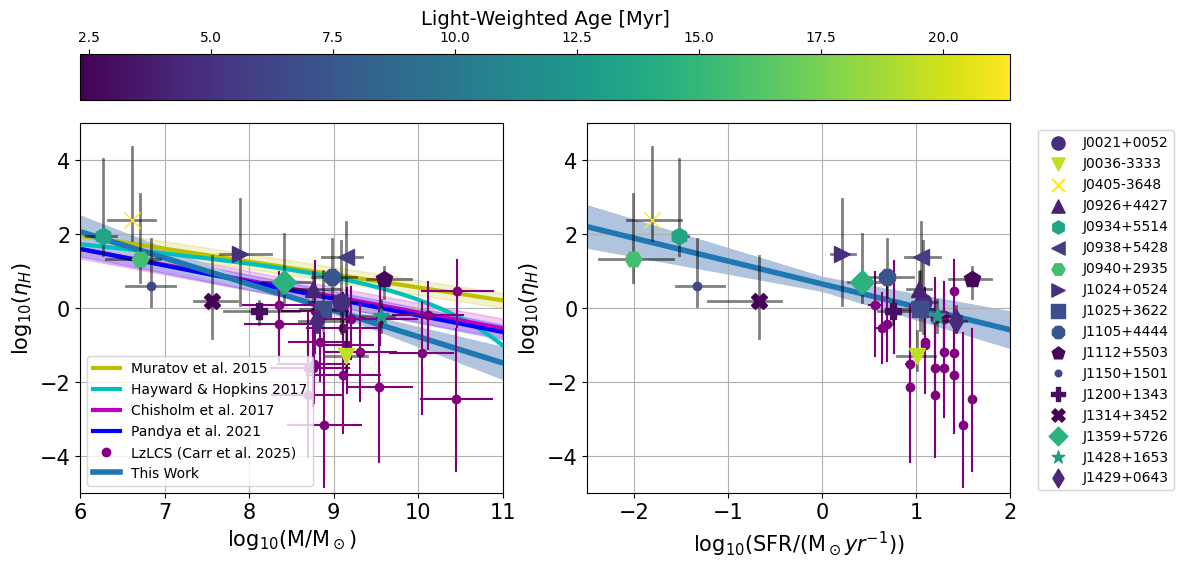}
    \includegraphics[width=0.9\linewidth]{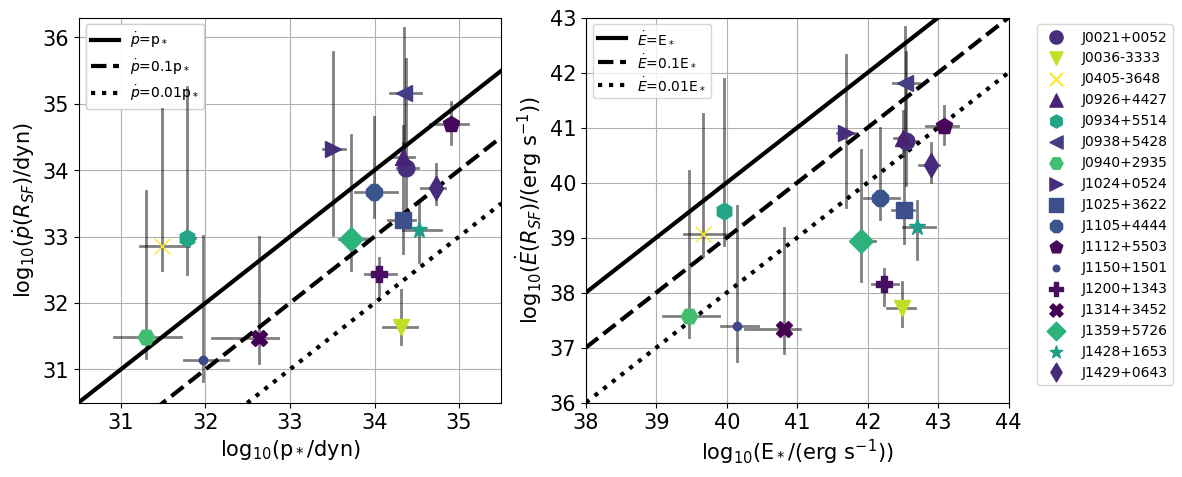}
    \includegraphics[width=0.9\linewidth]{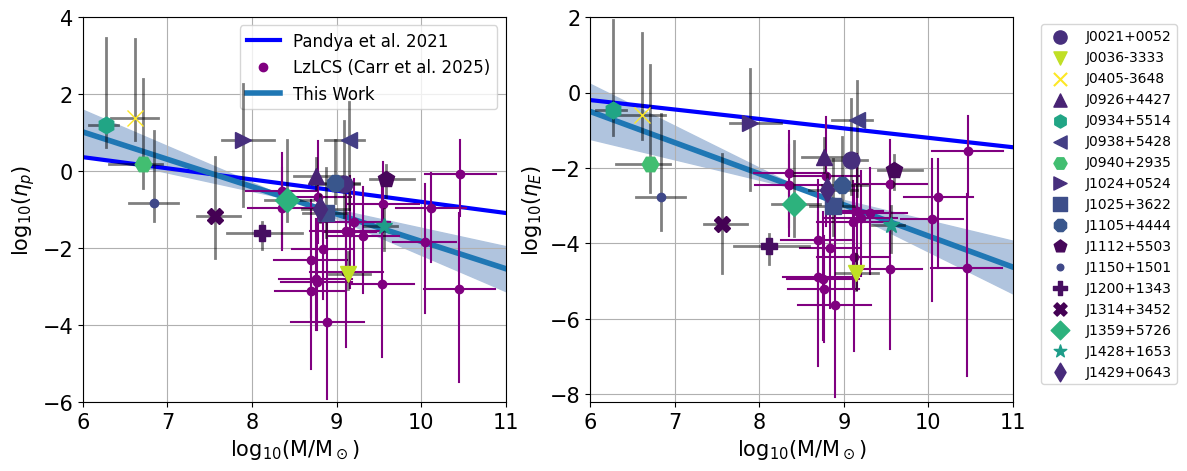}
    \caption{Upper left (Upper right): Comparison between $\eta_{H}$ and the stellar mass (SFR) for CLASSY galaxies (black) and LzLCS galaxies \citep[purple,][]{carr2025}. A few simulated models established in previous works are included as a reference, which follow the same trends as CLASSY and LzLCS. Center left: Comparison between $\dot p$ and $p_*$ as derived from the SFRs. Most galaxies lie within an order of magnitude of the one-to-one line, implying an efficient transfer of momentum in these CLASSY galaxies. Center right: Comparison between $\dot E$ and $E_*$ as derived from the SFRs. Galaxies lie systematically below the one-to-one line implies there does not exist an efficient transfer of energy in most of these CLASSY galaxies. Lower left (Lower right): Comparison between $\eta_{p}$ ($\eta_{E}$) and stellar mass. We also plot predictions from the Fire-2 simulations \citep{pandya2021}. The momentum loading factors from CLASSY and LzLCS are similar in magnitude to the Fire-2 simulations. However, there is a larger discrepancy between CLASSY and LzLCS and the energy loading factor, with the CLASSY and LzLCS measurements typically several orders of magnitude smaller. The color of the CLASSY points in all panels in the light-weighted age.}
    \label{fig:eta}
\end{figure*}

We now turn our attention to the related mass loading factor. 
As star formation results in the deposition of momentum and energy into the gas surrounding the star-forming regions, SFR can be used to normalize the mass outflow rates. 
This normalized value is typically referred to as the mass loading factor, $\eta_H=\dot M (R_{SF})/SFR$, and provides a measurement of how efficient star-formation is in driving outflows. 
In Figure \ref{fig:eta} we show the mass loading factor as a function of the galaxies' stellar mass in the upper left panel. We include several observational and theoretical expectations \citep{muratov2015,hayward2017,chisholm2017,pandya2021}. 
We also include points (in purple) from the LzLCS survey. \citet{carr2025} measured the loading factors from a subset of 29 LzLCS \mgii\ lines using SALT, in a very similar manner as this work. These sources are more massive than the CLASSY sample, allowing us to investigate how the loading factor changes at even larger masses. 
The mass loading factor typically decreases with stellar mass in the CLASSY sample. The more-massive LzLCS galaxies agree with this trend, with the LzLCS sample typically having lower loading factors than CLASSY. 
Compared with the plotted observational and theoretical expectations, we find that the magnitude of mass loading factors for the CLASSY galaxies tend to be in fairly good agreement with these models for a given stellar mass. This could be because a larger stellar mass causes a denser ISM and more gas recycling, decreasing the efficiency of such flows. Additionally, there may be an observational bias, as the hottest gas, which is more dominant in more massive sources, is not traced by the UV silicon lines \citep{pandya2021}.
The upper right panel of Figure~\ref{fig:eta} which shows the comparison between the mass loading factor and the star-formation rate\footnote{It is also worth noting that as $\eta$ is derived from the SFR, a degeneracy may have been introduced \citep[see][]{heckman2015}. }. We find as the SFR increases, $\eta_H$ decreases. This may simply be an extension of existing stellar mass/SFR correlations, but may also indicate that in highly star-forming sources, there are larger quantities of very hot gas, which is not traced by the UV silicon lines.

\subsection{Momentum and Energy Outflow Rates}\label{sec:pande}
We now compute the momentum and kinetic energy outflow rates for the CLASSY galaxies. The momentum outflow rate (the momentum flux carried through the outflow) is given by $\dot p_{Si,ion}=\dot M_{Si,ion} v$ whereas the kinetic energy outflow rate (the kinetic energy flux carried through the outflow) is given by $\dot E_{Si,ion}=\frac{1}{2}\dot M_{Si,ion} v^2$ \citep{chisholm2017}. We measure these rates at $v=v(R_{SF})=v_0$ since $\dot M_{Si,ion}$ is also computed at $R_{SF}$.
In the same manner as the computation for the mass outflow rate, the momentum and kinetic energy outflow rates are computed individually for each ionization state of silicon, summed together to get the total outflow rates of silicon, and then scaled to the hydrogen outflow rates. The kinetic energy outflow rate is a particularly simplified expression of energy outflow, as it does not include thermal, turbulent, or magnetic components. Similar to the mass outflow rate, the momentum and kinetic energy outflow rates can be normalized by the deposition of momentum and energy from star-formation. 
We take the momentum deposition rate ($p_{*}$), and the energy deposition rates ($E_{*}$) as those presented in \citet{murray2005} and \citet{xu2022,leitherer1999} respectively, where the momentum and energy deposition rates from supernovae are given by $p_{*} \approx 2\times 10^{33} \left ( \frac{SFR}{M_\odot yr^{-1}} \right )$dynes and $E_{*} \approx 3\times 10^{41} \left ( \frac{SFR}{M_\odot yr^{-1}} \right ) \rm erg\ s^{-1}$. As a side note, these relations only include direct momentum and energy deposition from supernovae, and would likely increase with the inclusion of other factors such as stellar winds, and therefore these may be conservative estimates. We note that some sources are characterized by very young light-weighted ages and may therefore be observed before the onset of supernova activity. In such cases, the available mechanical energy may actually be overestimated because only stellar winds contribute to feedback \citep{fierlinger2016,zhang2018}, with an energy input predicted to be approximately $\sim10\%$ of that supplied by supernovae, \citep{leithererheckman1995,saldanalopez2026}, although this is also a strong function of stellar metallicity \citep{vink2001,jecmenoey2023}. However, disentangling the relative contributions of stellar winds and supernovae to the driving of galactic outflows remains challenging \citep[see][]{saldanalopez2026}, in particular given that stars from outside the observing aperture may contribute to the outflow as well. Additionally, the light-weighted ages may also overlook older starbursts and their associated supernovae that may contribute to drive the observed outflow. Therefore following \citet{xu2022}, we retain these definitions of the energy-deposition rates as a standard basis for comparison, while keeping these caveats in mind.

In Figure~\ref{fig:eta}, we show the comparisons between $\dot p$ and $p_*$ (center left panel) and between $\dot E$ and $E_*$ (center right panel). We include the light-weight ages in each panel: for the youngest sources, $p_{*}$ and $E_{*}$ may be overestimated, moving these sources left on both center panels.
The momentum deposition rate provided by the starbursts observed in the CLASSY galaxies is consistent in magnitude with the observed momentum outflow rate for a majority of CLASSY galaxies. With the aforementioned caveats in mind, a majority of the sample agrees in magnitude between the supernovae deposited momentum flux and the measured momentum flux in the outflow.
In contrast, the kinetic energy deposition rate is typically $\sim2$ dex larger than the estimated kinetic energy outflow rates, indicating kinetic energy is lost within the outflow. This may suggest that momentum is transferred more efficiently than kinetic energy in these systems. \citet{xu2022} was able to have some success in explaining the observed CLASSY outflows with a purely momentum-driven simple analytical model.
In the momentum driven regime, the outflowing wind has been predicted to radiatively cool as a function of radius, becoming an isothermal front with radiative pressure that deposits its momentum into the ISM \citep{costa2014}. In this scenario where radiative cooling is significant, momentum transferred to the gas from ram pressure leads to a gentler acceleration of the gas which allows for cold gas to reach large velocities prior to being destroyed \citep{spilker2020}. This ties together well with the measurements of large velocities observed in the \sitwo-probed cooler gas, compared to the \sithree\ and \sifour\ probes (see section~\ref{sec:propertycomparison}). However, we cannot exclude a model where the outflow is predominantly energy driven with a weak coupling efficiency and the momentum flux increases as the wind moves out due to mass loading \citep[e.g.,][]{fauchergiguere2012}. 
Therefore, even if the CLASSY outflows demonstrate an efficient transfer of momentum, we cannot definitively conclude that the outflows are purely momentum-driven in nature, in particular given that the hottest phase of the outflows are predicted to carry substantial mechanical energy relative to the cooler phases \citep{pandya2021} like those traced by the UV silicon lines. A mix of momentum-driven and thermalized/energy-driven regimes is likely more appropriate \citep[e.g.,][]{fluetsch2021}.

We now define the momentum loading factor as $\eta_p=\dot p/p_{*}$ and the energy loading factor as $\eta_E=\dot E/E_{*}$.
In Figure~\ref{fig:eta} we show the dependency of the momentum loading factors (lower left panel) and energy loading factors (lower right panel) with stellar mass. These figures also show the predictions from the FIRE-2 hydrodynamical simulations of \citet{pandya2021} (where the energy loading factor from \citet{pandya2021} is the total energy, not just kinetic energy). We again show the corresponding measurement for the typically more massive LzLCS galaxies in purple.

\begin{figure*}
    \centering
    \includegraphics[width=0.99\linewidth]{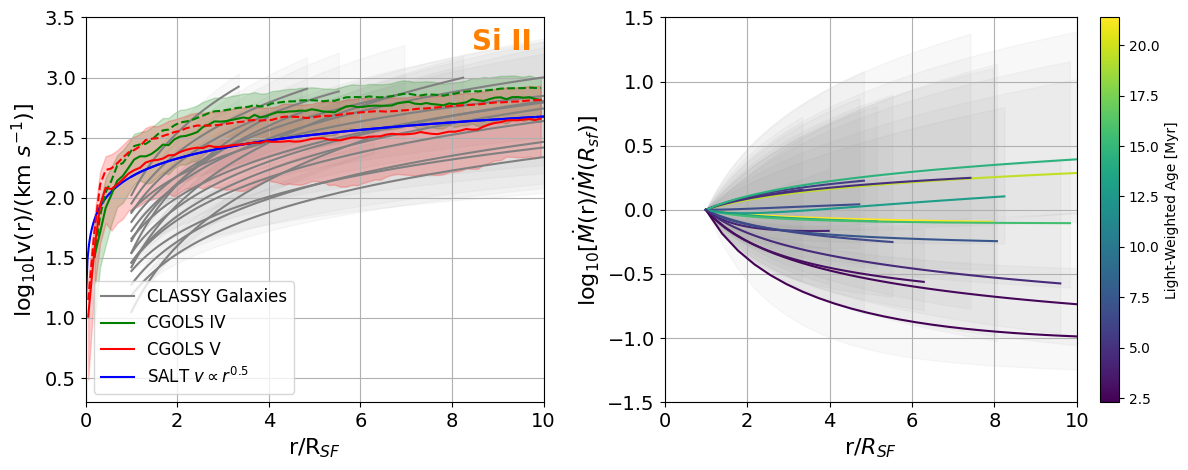}
    \caption{Left panel: Radial profiles of the velocity of the CLASSY outflows in the \sitwo\ traced cool gas. The CLASSY outflows are predicted to rise faster than the CGOLS simulations predict, but have similar terminal velocities as the CGOLS simulations. The dashed (solid) lines indicated the mean (median) value of the distribution of gas at a specific radius in the CGOLS outflows. CLASSY outflows terminate at their respective $r=R_W$. The blue curve shows a mock velocity profile within the SALT framework that well described the CGOLS V's $v(r)$. Right panel: Radial profiles of the hydrogen mass outflow rate of the CLASSY outflows. The CLASSY galaxies show a diversity of systems, with most outflow rates decreasing with radius, while others actually increase, often those with older underlying populations. Outflows terminate at their respective $r=R_W$.}
    \label{fig:radialmor}
\end{figure*}

We find that the momentum loading factor has an observable dependency on a galaxy's stellar mass, in agreement with the results of the FIRE-2 simulations: more massive galaxies tend to have lower momentum loading, most evident when comparing the typically more massive LzLCS galaxies with the less massive CLASSY ones. We find that light-weighted age does not play a huge role in the loading factors (in particular relative to mass). A similar trend is seen with the kinetic energy loading factor and the stellar masses.
However, we find a strong discrepancy between the magnitude of the observed kinetic energy loading factors of both CLASSY and LzLCS and the total energy as predicted from the FIRE-2 simulations. The CLASSY energy loading factors are up to three dex smaller, while the LzLCS ones are as much as five dex. Given the sharper slope of the line of best-fit for the CLASSY galaxies in the bottom right panel of figure~\ref{fig:eta}, we find that the kinetic energy loading factor makes up a larger proportion of the predicted total energy loading factor from FIRE-2 in less massive sources. One potential explanation for this observation is that since less massive sources have weaker potential wells, less kinetic energy is lost when driving the outflows.

\subsection{Radial Profiles of the Outflows}\label{sec:radial}
In the previous sections, we investigated outflow properties at $R_{SF}$. However, we know that outflow properties vary with radius. 
By assumption, the SALT model defines the velocity and density profiles as power laws (with indices $\gamma$ and $\delta$ respectively). Testing on independent simulations have shown $\gamma$ and $\delta$ are more weakly constrained relative to the other SALT parameters \citep[see][]{carr2023}, although \citet{carrsmith2025} found they were more strongly recovered in highly blue-shifted spectra, which is the case for a majority of the 17 CLASSY outflows. 
Nevertheless, \citet{carr2023} showed that even with weaker constraints of these parameters, the radially variable properties including the mass outflow rate measured with SALT were still well constrained at various radii (including at both $R_{SF}$ and $R_W$).

In the left panel of Figure~\ref{fig:radialmor}, we show the velocity of the cool outflows traced by \sitwo\ as a function of radius for the 17 CLASSY galaxies in gray (with the uncertainty shaded in light gray). Each outflow terminates at its respective $R_W$. 
We also plot the cool phase outflows from the Cholla Galactic OutfLow Simulations (CGOLS) of \citet{schneider2020} (CGOLS IV, green) and \citet{schneider2024} (CGOLS V, red). 
CGOLS is a set of three-dimensional, hydrodynamical global galaxy simulations and are known for being highly resolved numerical simulations of galaxy outflows (with a resolution of $\approx5$ pc) \citep{schneider2018}. 
Initialized from a rotating gas disk surrounded by a static halo (in hydrostatic equilibrium with the gravitational potential), mass and thermal energy are injecting into the center of the mock galaxy, driving a biconical collimated wind. In CGOLS IV, mass and energy are deposited into randomly distributed ``stellar" clusters within the central portions of the galaxy, which are turned on and off, mimicking bursts of star-formation. In CGOLS V, this was modified with the addition of a cluster mass function, in which the clusters where given different masses distributed in the same manner as nearby galaxies including M82, a galaxy hosting an outflow that CGOLS has been able to recreate observations of quite successfully \citep{schneider2018}. 
Thus, CGOLS can act as a standard of comparison for multiphase outflows initiated by star-formation, with the cool phase gas being the most appropriate proxy for \sitwo\ traced gas. The dashed (solid) lines in the left panel of Figure~\ref{fig:radialmor} for the CGOLS simulations represent the mean (median) value of the distribution of cool gas at each radius in the CGOLS outflows. We normalize the radii of the outflows to create a standard of comparison given the diversity of systems. The CGOLS simulations have a cluster seeding radius of $1$ kpc \citep{schneider2020}, which we treat as a proxy for our definition of $R_{SF}$. 

The SALT framework from \citetalias{huberty2024} can reasonably well trace the $v(r)$ of the CGOLS models:
we show an example of such with the blue line in left panel of figure~\ref{fig:radialmor}, where $v(r)=150\rm\ km\ s^{-1} (\frac{r}{RSF})^{0.5}$  can well describe the median $v(r)$ from CGOLS V (we also plot this curve at $r<R_{SF}$ to show the continued similiarity at lower radii).
However in the observed CLASSY outflows, at $R_{SF}$, the outflows are always predicted to have a lower initial velocity than the seeding radius of CGOLS. Additionally, the velocity of the CLASSY outflows is always measured to increase more rapidly with radius at $r>R_{SF}$ than the CGOLS outflows, with an average power law index of $\gamma\sim1.2$. The median (mean) CGOLS IV outflows level out near $650\ \rm km\ s^{-1}$ ($815\ \rm km\ s^{-1}$). The median (mean) CGOLS V outflows level out near $450\ \rm km\ s^{-1}$ ($650\ \rm km\ s^{-1}$). The average CLASSY outflow reached a maximum velocity at $R_W$ of $\sim620\ \rm km\ s^{-1}$. Thus we find that the maximum velocity of the CGOLS and CLASSY outflows are typically in strong agreement, even if the initial velocities and velocities gradients differ. Interestingly, \citet{xu2026} independently derived outflow velocities for a subset of CLASSY galaxies with the PCM model that also quantitatively agrees with the SALT-measured velocity evolution of the outflow with radius, in which the maximum velocity of the outflows varies between $100-1000\ \rm km\ s^{-1}$, with the typical value being around $400-500\ \rm km\ s^{-1}$, which is logical as the PCM tends to measure slightly radially smaller outflows than SALT. 
The velocity discrepancy between CGOLS and CLASSY at low radius (and thus at low velocity) may be due to the difficulty in disentangling the CLASSY outflow component from the static ISM or superbubble components. 
Regardless, this comparison demonstrates that the observed velocity evolution measured in CLASSY galaxies is not unreasonable relative to hydrodynamic CGOLS simulations, and predicts similar maximum velocities.

We can use the velocity profiles, alongside corresponding density profiles to obtain the hydrogen mass outflow rate as a function of radius relative to the mass outflow rate at $R_{SF}$, shown in the right panel of Figure~\ref{fig:radialmor}. We color the outflow rates by the light-weighted ages from \citet{parker2025}. The shaded light gray areas indicate the uncertainty on each outflow. While the mass outflow rate uncertainties are large, \citet{carr2023} showed that SALT-derived $\dot M(R_W)$ is more likely to overestimate than underestimate the true value. We find that the galaxies with the oldest stellar-age populations are more likely to increase with radius (or remain roughly constant with radius), while the younger sources are more likely to decrease with radius. 

To speculate on the physical origin of these findings, one possibility is that in the youngest stellar populations (where the mass outflow rate decreases with radius), outflows are likely driven primarily by stellar winds, as the most massive stars have not yet had time to explode. The declining mass outflow with radius may indicate the presence of galactic fountains, in which re-accreting gas counteracts the outflow \citep{carr2022}. Such gas inflow would also naturally explain the very recent star formation.
In contrast, for sources where mass outflow rates increase with radius, this behavior could be explained by the rapid entrainment of cool gas with radius swept up by the outflow.
This is particularly true in the older stellar populations ($>10\rm\ Myr$) when supernova have fully turned on, and where mass outflow rates tend to increase or remain constant with radius. As noted in Section~\ref{sec:makeup}, these older populations are more likely to be more heavily composed of cool gas, which may be entrained from the local ISM.

\section{Conclusion}\label{sec:conclusion}

In this work, we analyzed detailed radiation transfer modeling conducted in \citet{huberty2024} in order to understand the properties of the CLASSY galaxy outflows. 
We find that the outflow properties inferred from UV silicon absorption lines, which trace different gas phases, vary significantly between phases. In particular, the cooler \sitwo\ phase often shows spatial distributions and kinematic properties that differ from those of the warmer \sithree\ and \sifour\ phases of the outflow. These discrepancies demonstrate that both the cool and warm phases are essential for accurately characterizing overall outflow rates. 
We find mass outflow rates typically increase with SFR, and that the mass loading factor scales inversely with stellar mass, in line with various simulations. This trend extends to larger mass sources from the LzLCS survey. The same is true of momentum and energy loading, though kinetic energy loading observations tend to be several dex in magnitude below the theoretical total energy loading factors, which is most discrepant for the most massive sources.

We also find that there is a possible light-weighted age dependence on the composition and properties of the outflows: we find that outflows driven by younger ($<5\rm\ Myr$) light-weighted stellar populations are more likely to have a column density dominated by cooler gas, have a larger $\dot M(R_{SF})$, and an $\dot M(r)$ that decreases with radius. We find that outflows driven by older ($>10\rm\ Myr$) light-weight populations are more likely to be dominated by warmer gas, have a smaller $\dot M(R_{SF})$, and an $\dot M(r)$ that increase or remains roughly constant with radius. 

There is growing evidence that outflows change drastically with age \citep[e.g.,][]{hayes2023,carrsmith2025,carrnature2025}. While the light-weighted age can provide some insight into this change over time, the light-weighted age does not give you the full picture, as this age is more biased by younger stellar populations than older ones. Older populations of stars may have influenced the porosity and density of the ISM in the past, which in turn will affect the spatial distributions and kinematics of outflows in the present. Additionally, the HST/COS aperture is known to not fully encompass all stellar populations, which may contribute to additional uncertainties in the physics that drives these outflows.
While numerous simulations are able to reproduce the CLASSY observational measurements, persistent discrepancies suggest that significant gaps in our understanding remain.
The upcoming Habitable Worlds Observatory will be an important development in the study of outflows: its larger aperture compared to that of the Hubble Space Telescope will alleviate many of the restrictions imposed by Hubble's smaller aperture, and will allow for a more expansive look into the extended circumgalactic medium and the underlying associated stellar populations.

%\columnbreak
%\newpage
\section{Acknowledgement}
M. Huberty would like to thank Evan Schneider for access to the CGOLS data. 
We acknowledge the Minnesota Supercomputing Institute (MSI) at the University of Minnesota for providing the computational resources used in this project.
C. C. was supported by the NSFC grant W2433001 and the NSFC Talent-Introduction Program.  C. C. also acknowledges support from the University of Michigan through the ELT Fellowship Program. 
The data presented in this article were obtained from the Mikulski Archive for Space Telescopes (MAST) at the Space Telescope Science Institute. The specific observations analyzed can be accessed via the CLASSY HLSP \dataset[doi: 10.17909/m3fq-jj25]{https://doi.org/10.17909/m3fq-jj25}.

\software{NumPy \citep{Numpy:2020}; SciPy \citep{Scipy:2020}; AstroPy \citep{Astropy:2013,Astropy:2018,Astropy:2022}; Matplotlib \citep{Matplotlib:2007}.}

\clearpage
%\bibliography{refr22}
%\begin{thebibliography}{}

%\end{thebibliography}

%\printbibliography

\bibliography{refr}{}

@ARTICLE{vink2001,
       author = {{Vink}, Jorick S. and {de Koter}, A. and {Lamers}, H.~J.~G.~L.~M.},
        title = "{Mass-loss predictions for O and B stars as a function of metallicity}",
      journal = {\aap},
         year = 2001,
        month = apr,
       volume = {369},
        pages = {574-588},
          doi = {10.1051/0004-6361:20010127},
archivePrefix = {arXiv},
       eprint = {astro-ph/0101509},
 primaryClass = {astro-ph},
       adsurl = {https://ui.adsabs.harvard.edu/abs/2001A&A...369..574V}
}

@ARTICLE{jecmenoey2023,
       author = {{Jecmen}, Michelle C. and {Oey}, M.~S.},
        title = "{Delayed Massive-star Mechanical Feedback at Low Metallicity}",
      journal = {\apj},
         year = 2023,
        month = dec,
       volume = {958},
       number = {2},
          eid = {149},
        pages = {149},
          doi = {10.3847/1538-4357/ad0460},
archivePrefix = {arXiv},
       eprint = {2310.10589},
 primaryClass = {astro-ph.GA},
       adsurl = {https://ui.adsabs.harvard.edu/abs/2023ApJ...958..149J}
}

@ARTICLE{Numpy:2020,
         title = "{Array programming with {NumPy}}",
        author = {Charles R. Harris and K. Jarrod Millman and St{\'{e}}fan J. van der Walt and Ralf Gommers and Pauli Virtanen and David Cournapeau and Eric Wieser and Julian Taylor and Sebastian Berg and Nathaniel J. Smith and Robert Kern and Matti Picus and Stephan Hoyer and Marten H. van Kerkwijk and Matthew Brett and Allan Haldane and Jaime Fern{\'{a}}ndez del R{\'{i}}o and Mark Wiebe and Pearu Peterson and Pierre G{\'{e}}rard-Marchant and Kevin Sheppard and Tyler Reddy and Warren Weckesser and Hameer Abbasi and Christoph Gohlke and Travis E. Oliphant},
          year = 2020,
         month = sep,
       journal = {Nature},
        volume = {585},
        number = {7825},
         pages = {357--362},
           doi = {10.1038/s41586-020-2649-2},
     publisher = {Springer Science and Business Media {LLC}},
           url = {https://doi.org/10.1038/s41586-020-2649-2}
}

@ARTICLE{Scipy:2020,
  author  = {Virtanen, Pauli and Gommers, Ralf and Oliphant, Travis E. and Haberland, Matt and Reddy, Tyler and Cournapeau, David and Burovski, Evgeni and Peterson, Pearu and Weckesser, Warren and Bright, Jonathan and {van der Walt}, St{\'e}fan J. and Brett, Matthew and Wilson, Joshua and Millman, K. Jarrod and Mayorov, Nikolay and Nelson, Andrew R. J. and Jones, Eric and Kern, Robert and Larson, Eric and Carey, C J and Polat, {\.I}lhan and Feng, Yu and Moore, Eric W. and {VanderPlas}, Jake and Laxalde, Denis and Perktold, Josef and Cimrman, Robert and Henriksen, Ian and Quintero, E. A. and Harris, Charles R. and Archibald, Anne M. and Ribeiro, Ant{\^o}nio H. and Pedregosa, Fabian and {van Mulbregt}, Paul and {SciPy 1.0 Contributors}},
  title   = "{{{SciPy} 1.0: Fundamental Algorithms for Scientific Computing in Python}}",
  journal = {Nature Methods},
  year    = 2020,
  volume  = {17},
  pages   = {261--272},
  adsurl  = {https://rdcu.be/b08Wh},
  doi     = {10.1038/s41592-019-0686-2},
}

@ARTICLE{Matplotlib:2007,
  Author    = {Hunter, J. D.},
  Title     = {Matplotlib: A 2D graphics environment},
  Journal   = {Computing in Science \& Engineering},
  Volume    = {9},
  Number    = {3},
  Pages     = {90--95},
  publisher = {IEEE COMPUTER SOC},
  doi       = {10.1109/MCSE.2007.55},
  year      = 2007
}

@ARTICLE{Astropy:2013,
         Adsurl = {http://adsabs.harvard.edu/abs/2013A%26A...558A..33A},
  Archiveprefix = {arXiv},
         Author = {{Astropy Collaboration} and {Robitaille}, T.~P. and {Tollerud}, E.~J. and {Greenfield}, P. and {Droettboom}, M. and {Bray}, E. and {Aldcroft}, T. and {Davis}, M. and {Ginsburg}, A. and {Price-Whelan}, A.~M. and {Kerzendorf}, W.~E. and {Conley}, A. and {Crighton}, N. and {Barbary}, K. and {Muna}, D. and {Ferguson}, H. and {Grollier}, F. and {Parikh}, M.~M. and {Nair}, P.~H. and {Unther}, H.~M. and {Deil}, C. and {Woillez}, J. and {Conseil}, S. and {Kramer}, R. and {Turner}, J.~E.~H. and {Singer}, L. and {Fox}, R. and {Weaver}, B.~A. and {Zabalza}, V. and {Edwards}, Z.~I. and {Azalee Bostroem}, K. and {Burke}, D.~J. and {Casey}, A.~R. and {Crawford}, S.~M. and {Dencheva}, N. and {Ely}, J. and {Jenness}, T. and {Labrie}, K. and {Lim}, P.~L. and {Pierfederici}, F. and {Pontzen}, A. and {Ptak}, A. and {Refsdal}, B. and {Servillat}, M. and {Streicher}, O.},
            Doi = {10.1051/0004-6361/201322068},
            Eid = {A33},
         Eprint = {1307.6212},
        Journal = {\aap},
          Month = oct,
          Pages = {A33},
   Primaryclass = {astro-ph.IM},
          Title = {{Astropy: A community Python package for astronomy}},
         Volume = 558,
           Year = 2013,
            url = {https://dx.doi.org/10.1051/0004-6361/201322068}
}

@ARTICLE{Astropy:2018,
       author = {{Astropy Collaboration} and {Price-Whelan}, A.~M. and {Sip{\H{o}}cz}, B.~M. and {G{\"u}nther}, H.~M. and {Lim}, P.~L. and {Crawford}, S.~M. and {Conseil}, S. and {Shupe}, D.~L. and {Craig}, M.~W. and {Dencheva}, N. and {Ginsburg}, A. and {VanderPlas}, J.~T. and {Bradley}, L.~D. and {P{\'e}rez-Su{\'a}rez}, D. and {de Val-Borro}, M. and {Aldcroft}, T.~L. and {Cruz}, K.~L. and {Robitaille}, T.~P. and {Tollerud}, E.~J. and {Ardelean}, C. and {Babej}, T. and {Bach}, Y.~P. and {Bachetti}, M. and {Bakanov}, A.~V. and {Bamford}, S.~P. and {Barentsen}, G. and {Barmby}, P. and {Baumbach}, A. and {Berry}, K.~L. and {Biscani}, F. and {Boquien}, M. and {Bostroem}, K.~A. and {Bouma}, L.~G. and {Brammer}, G.~B. and {Bray}, E.~M. and {Breytenbach}, H. and {Buddelmeijer}, H. and {Burke}, D.~J. and {Calderone}, G. and {Cano Rodr{\'\i}guez}, J.~L. and {Cara}, M. and {Cardoso}, J.~V.~M. and {Cheedella}, S. and {Copin}, Y. and {Corrales}, L. and {Crichton}, D. and {D'Avella}, D. and {Deil}, C. and {Depagne}, {\'E}. and {Dietrich}, J.~P. and {Donath}, A. and {Droettboom}, M. and {Earl}, N. and {Erben}, T. and {Fabbro}, S. and {Ferreira}, L.~A. and {Finethy}, T. and {Fox}, R.~T. and {Garrison}, L.~H. and {Gibbons}, S.~L.~J. and {Goldstein}, D.~A. and {Gommers}, R. and {Greco}, J.~P. and {Greenfield}, P. and {Groener}, A.~M. and {Grollier}, F. and {Hagen}, A. and {Hirst}, P. and {Homeier}, D. and {Horton}, A.~J. and {Hosseinzadeh}, G. and {Hu}, L. and {Hunkeler}, J.~S. and {Ivezi{\'c}}, {\v{Z}}. and {Jain}, A. and {Jenness}, T. and {Kanarek}, G. and {Kendrew}, S. and {Kern}, N.~S. and {Kerzendorf}, W.~E. and {Khvalko}, A. and {King}, J. and {Kirkby}, D. and {Kulkarni}, A.~M. and {Kumar}, A. and {Lee}, A. and {Lenz}, D. and {Littlefair}, S.~P. and {Ma}, Z. and {Macleod}, D.~M. and {Mastropietro}, M. and {McCully}, C. and {Montagnac}, S. and {Morris}, B.~M. and {Mueller}, M. and {Mumford}, S.~J. and {Muna}, D. and {Murphy}, N.~A. and {Nelson}, S. and {Nguyen}, G.~H. and {Ninan}, J.~P. and {N{\"o}the}, M. and {Ogaz}, S. and {Oh}, S. and {Parejko}, J.~K. and {Parley}, N. and {Pascual}, S. and {Patil}, R. and {Patil}, A.~A. and {Plunkett}, A.~L. and {Prochaska}, J.~X. and {Rastogi}, T. and {Reddy Janga}, V. and {Sabater}, J. and {Sakurikar}, P. and {Seifert}, M. and {Sherbert}, L.~E. and {Sherwood-Taylor}, H. and {Shih}, A.~Y. and {Sick}, J. and {Silbiger}, M.~T. and {Singanamalla}, S. and {Singer}, L.~P. and {Sladen}, P.~H. and {Sooley}, K.~A. and {Sornarajah}, S. and {Streicher}, O. and {Teuben}, P. and {Thomas}, S.~W. and {Tremblay}, G.~R. and {Turner}, J.~E.~H. and {Terr{\'o}n}, V. and {van Kerkwijk}, M.~H. and {de la Vega}, A. and {Watkins}, L.~L. and {Weaver}, B.~A. and {Whitmore}, J.~B. and {Woillez}, J. and {Zabalza}, V. and {Astropy Contributors}},
        title = "{The Astropy Project: Building an Open-science Project and Status of the v2.0 Core Package}",
      journal = {\aj},
         year = 2018,
        month = sep,
       volume = {156},
       number = {3},
          eid = {123},
        pages = {123},
          doi = {10.3847/1538-3881/aabc4f},
archivePrefix = {arXiv},
       eprint = {1801.02634},
 primaryClass = {astro-ph.IM},
       adsurl = {https://ui.adsabs.harvard.edu/abs/2018AJ....156..123A}
}

@ARTICLE{Astropy:2022,
       author = {{Astropy Collaboration} and {Price-Whelan}, Adrian M. and {Lim}, Pey Lian and {Earl}, Nicholas and {Starkman}, Nathaniel and {Bradley}, Larry and {Shupe}, David L. and {Patil}, Aarya A. and {Corrales}, Lia and {Brasseur}, C.~E. and {N{"o}the}, Maximilian and {Donath}, Axel and {Tollerud}, Erik and {Morris}, Brett M. and {Ginsburg}, Adam and {Vaher}, Eero and {Weaver}, Benjamin A. and {Tocknell}, James and {Jamieson}, William and {van Kerkwijk}, Marten H. and {Robitaille}, Thomas P. and {Merry}, Bruce and {Bachetti}, Matteo and {G{"u}nther}, H. Moritz and {Aldcroft}, Thomas L. and {Alvarado-Montes}, Jaime A. and {Archibald}, Anne M. and {B{'o}di}, Attila and {Bapat}, Shreyas and {Barentsen}, Geert and {Baz{'a}n}, Juanjo and {Biswas}, Manish and {Boquien}, M{'e}d{'e}ric and {Burke}, D.~J. and {Cara}, Daria and {Cara}, Mihai and {Conroy}, Kyle E. and {Conseil}, Simon and {Craig}, Matthew W. and {Cross}, Robert M. and {Cruz}, Kelle L. and {D'Eugenio}, Francesco and {Dencheva}, Nadia and {Devillepoix}, Hadrien A.~R. and {Dietrich}, J{"o}rg P. and {Eigenbrot}, Arthur Davis and {Erben}, Thomas and {Ferreira}, Leonardo and {Foreman-Mackey}, Daniel and {Fox}, Ryan and {Freij}, Nabil and {Garg}, Suyog and {Geda}, Robel and {Glattly}, Lauren and {Gondhalekar}, Yash and {Gordon}, Karl D. and {Grant}, David and {Greenfield}, Perry and {Groener}, Austen M. and {Guest}, Steve and {Gurovich}, Sebastian and {Handberg}, Rasmus and {Hart}, Akeem and {Hatfield-Dodds}, Zac and {Homeier}, Derek and {Hosseinzadeh}, Griffin and {Jenness}, Tim and {Jones}, Craig K. and {Joseph}, Prajwel and {Kalmbach}, J. Bryce and {Karamehmetoglu}, Emir and {Ka{l}uszy{'n}ski}, Miko{l}aj and {Kelley}, Michael S.~P. and {Kern}, Nicholas and {Kerzendorf}, Wolfgang E. and {Koch}, Eric W. and {Kulumani}, Shankar and {Lee}, Antony and {Ly}, Chun and {Ma}, Zhiyuan and {MacBride}, Conor and {Maljaars}, Jakob M. and {Muna}, Demitri and {Murphy}, N.~A. and {Norman}, Henrik and {O'Steen}, Richard and {Oman}, Kyle A. and {Pacifici}, Camilla and {Pascual}, Sergio and {Pascual-Granado}, J. and {Patil}, Rohit R. and {Perren}, Gabriel I. and {Pickering}, Timothy E. and {Rastogi}, Tanuj and {Roulston}, Benjamin R. and {Ryan}, Daniel F. and {Rykoff}, Eli S. and {Sabater}, Jose and {Sakurikar}, Parikshit and {Salgado}, Jes{'u}s and {Sanghi}, Aniket and {Saunders}, Nicholas and {Savchenko}, Volodymyr and {Schwardt}, Ludwig and {Seifert-Eckert}, Michael and {Shih}, Albert Y. and {Jain}, Anany Shrey and {Shukla}, Gyanendra and {Sick}, Jonathan and {Simpson}, Chris and {Singanamalla}, Sudheesh and {Singer}, Leo P. and {Singhal}, Jaladh and {Sinha}, Manodeep and {Sip{H{o}}cz}, Brigitta M. and {Spitler}, Lee R. and {Stansby}, David and {Streicher}, Ole and {{{S}}umak}, Jani and {Swinbank}, John D. and {Taranu}, Dan S. and {Tewary}, Nikita and {Tremblay}, Grant R. and {Val-Borro}, Miguel de and {Van Kooten}, Samuel J. and {Vasovi{'c}}, Zlatan and {Verma}, Shresth and {de Miranda Cardoso}, Jos{'e} Vin{'i}cius and {Williams}, Peter K.~G. and {Wilson}, Tom J. and {Winkel}, Benjamin and {Wood-Vasey}, W.~M. and {Xue}, Rui and {Yoachim}, Peter and {Zhang}, Chen and {Zonca}, Andrea and {Astropy Project Contributors}},
        title = "{The Astropy Project: Sustaining and Growing a Community-oriented Open-source Project and the Latest Major Release (v5.0) of the Core Package}",
      journal = {\apj},
         year = 2022,
        month = aug,
       volume = {935},
       number = {2},
          eid = {167},
        pages = {167},
          doi = {10.3847/1538-4357/ac7c74},
archivePrefix = {arXiv},
       eprint = {2206.14220},
 primaryClass = {astro-ph.IM},
       adsurl = {https://ui.adsabs.harvard.edu/abs/2022ApJ...935..167A}
}

@ARTICLE{xu2026,
       author = {{Xu}, Xinfeng and {Fielding}, Drummond and {Heckman}, Timothy and {Bryan}, Greg L. and {Henry}, Alaina and {Arellano-C{\'o}rdova}, Karla Z. and {Carr}, Cody and {Chisholm}, John and {Faucher-Gigu{\`e}re}, Claude-Andr{\'e} and {Hayes}, Matthew and {Huberty}, Mason and {Jennings}, Michael and {Martin}, Crystal L. and {Scarlata}, Claudia and {Strom}, Allison L.},
        title = "{Resolving the Unresolved Galactic Winds in Multiphase Models. I. Methodology and Application}",
      journal = {\apj},
         year = 2026,
        month = jun,
       volume = {1003},
       number = {2},
          eid = {155},
        pages = {155},
          doi = {10.3847/1538-4357/ae63d1},
archivePrefix = {arXiv},
       eprint = {2605.01105},
 primaryClass = {astro-ph.GA},
       adsurl = {https://ui.adsabs.harvard.edu/abs/2026ApJ..1003..155X}
}

@ARTICLE{schneider2024,
       author = {{Schneider}, Evan E. and {Mao}, S. Alwin},
        title = "{CGOLS V: Disk-wide Stellar Feedback and Observational Implications of the Cholla Galactic Wind Model}",
      journal = {\apj},
         year = 2024,
        month = may,
       volume = {966},
       number = {1},
          eid = {37},
        pages = {37},
          doi = {10.3847/1538-4357/ad2e8a},
archivePrefix = {arXiv},
       eprint = {2402.12474},
 primaryClass = {astro-ph.GA},
       adsurl = {https://ui.adsabs.harvard.edu/abs/2024ApJ...966...37S}
}

@ARTICLE{geen2016,
       author = {{Geen}, Sam and {Hennebelle}, Patrick and {Tremblin}, Pascal and {Rosdahl}, Joakim},
        title = "{Feedback in Clouds II: UV photoionization and the first supernova in a massive cloud}",
      journal = {\mnras},
         year = 2016,
        month = dec,
       volume = {463},
       number = {3},
        pages = {3129-3142},
          doi = {10.1093/mnras/stw2235},
archivePrefix = {arXiv},
       eprint = {1607.05487},
 primaryClass = {astro-ph.GA},
       adsurl = {https://ui.adsabs.harvard.edu/abs/2016MNRAS.463.3129G}
}

@ARTICLE{schneider2018,
       author = {{Schneider}, Evan E. and {Robertson}, Brant E.},
        title = "{Introducing CGOLS: The Cholla Galactic Outflow Simulation Suite}",
      journal = {\apj},
         year = 2018,
        month = jun,
       volume = {860},
       number = {2},
          eid = {135},
        pages = {135},
          doi = {10.3847/1538-4357/aac329},
archivePrefix = {arXiv},
       eprint = {1803.01008},
 primaryClass = {astro-ph.GA},
       adsurl = {https://ui.adsabs.harvard.edu/abs/2018ApJ...860..135S}
}

@ARTICLE{schneider2020,
       author = {{Schneider}, Evan E. and {Ostriker}, Eve C. and {Robertson}, Brant E. and {Thompson}, Todd A.},
        title = "{The Physical Nature of Starburst-driven Galactic Outflows}",
      journal = {\apj},
         year = 2020,
        month = may,
       volume = {895},
       number = {1},
          eid = {43},
        pages = {43},
          doi = {10.3847/1538-4357/ab8ae8},
archivePrefix = {arXiv},
       eprint = {2002.10468},
 primaryClass = {astro-ph.GA},
       adsurl = {https://ui.adsabs.harvard.edu/abs/2020ApJ...895...43S}
}

@ARTICLE{carr2023,
       author = {{Carr}, C. and {Michel-Dansac}, L. and {Blaizot}, J. and {Scarlata}, C. and {Henry}, A. and {Verhamme}, A.},
        title = "{Testing SALT Approximations with Numerical Radiation Transfer Code. I. Validity and Applicability}",
      journal = {\apj},
         year = 2023,
        month = jul,
       volume = {952},
       number = {1},
          eid = {88},
        pages = {88},
          doi = {10.3847/1538-4357/acd331},
archivePrefix = {arXiv},
       eprint = {2209.14473},
 primaryClass = {astro-ph.GA},
       adsurl = {https://ui.adsabs.harvard.edu/abs/2023ApJ...952...88C}
}

@ARTICLE{solar,
       author = {{Lodders}, Katharina},
        title = "{Relative Atomic Solar System Abundances, Mass Fractions, and Atomic Masses of the Elements and Their Isotopes, Composition of the Solar Photosphere, and Compositions of the Major Chondritic Meteorite Groups}",
      journal = {\ssr},
         year = 2021,
        month = apr,
       volume = {217},
       number = {3},
          eid = {44},
        pages = {44},
          doi = {10.1007/s11214-021-00825-8},
       adsurl = {https://ui.adsabs.harvard.edu/abs/2021SSRv..217...44L}
}

@ARTICLE{carr2022,
       author = {{Carr}, C. and {Scarlata}, C.},
        title = "{A Semianalytical Line Transfer Model. III. Galactic Inflows}",
      journal = {\apj},
         year = 2022,
        month = nov,
       volume = {939},
       number = {1},
          eid = {47},
        pages = {47},
          doi = {10.3847/1538-4357/ac93fa},
archivePrefix = {arXiv},
       eprint = {2209.14485},
 primaryClass = {astro-ph.GA},
       adsurl = {https://ui.adsabs.harvard.edu/abs/2022ApJ...939...47C}
}

@ARTICLE{carr2018,
       author = {{Carr}, Cody and {Scarlata}, Claudia and {Panagia}, Nino and {Henry}, Alaina},
        title = "{A Semi-analytical Line Transfer (SALT) Model. II: The Effects of a Bi-conical Geometry}",
      journal = {\apj},
         year = 2018,
        month = jun,
       volume = {860},
       number = {2},
          eid = {143},
        pages = {143},
          doi = {10.3847/1538-4357/aac48e},
archivePrefix = {arXiv},
       eprint = {1805.05981},
 primaryClass = {astro-ph.GA},
       adsurl = {https://ui.adsabs.harvard.edu/abs/2018ApJ...860..143C}
}

@ARTICLE{carr2021,
       author = {{Carr}, Cody and {Scarlata}, Claudia and {Henry}, Alaina and {Panagia}, Nino},
        title = "{The Effects of Biconical Outflows on Ly{\ensuremath{\alpha}} Escape from Green Peas}",
      journal = {\apj},
         year = 2021,
        month = jan,
       volume = {906},
       number = {2},
          eid = {104},
        pages = {104},
          doi = {10.3847/1538-4357/abc7c3},
archivePrefix = {arXiv},
       eprint = {2011.02549},
 primaryClass = {astro-ph.GA},
       adsurl = {https://ui.adsabs.harvard.edu/abs/2021ApJ...906..104C}
}

@ARTICLE{berg2022,
       author = {{Berg}, Danielle A. and {James}, Bethan L. and {King}, Teagan and {McDonald}, Meaghan and {Chen}, Zuyi and {Chisholm}, John and {Heckman}, Timothy and {Martin}, Crystal L. and {Stark}, Dan P. and {Aloisi}, Alessandra and {Amor{\'\i}n}, Ricardo O. and {Arellano-C{\'o}rdova}, Karla Z. and {Bayliss}, Matthew and {Bordoloi}, Rongmon and {Brinchmann}, Jarle and {Charlot}, St{\'e}phane and {Chevallard}, Jacopo and {Clark}, Ilyse and {Erb}, Dawn K. and {Feltre}, Anna and {Gronke}, Max and {Hayes}, Matthew and {Henry}, Alaina and {Hernandez}, Svea and {Jaskot}, Anne and {Jones}, Tucker and {Kewley}, Lisa J. and {Kumari}, Nimisha and {Leitherer}, Claus and {Llerena}, Mario and {Maseda}, Michael and {Mingozzi}, Matilde and {Nanayakkara}, Themiya and {Ouchi}, Masami and {Plat}, Adele and {Pogge}, Richard W. and {Ravindranath}, Swara and {Rigby}, Jane R. and {Sanders}, Ryan and {Scarlata}, Claudia and {Senchyna}, Peter and {Skillman}, Evan D. and {Steidel}, Charles C. and {Strom}, Allison L. and {Sugahara}, Yuma and {Wilkins}, Stephen M. and {Wofford}, Aida and {Xu}, Xinfeng and {Classy Team}},
        title = "{The COS Legacy Archive Spectroscopy Survey (CLASSY) Treasury Atlas}",
      journal = {\apjs},
         year = 2022,
        month = aug,
       volume = {261},
       number = {2},
          eid = {31},
        pages = {31},
          doi = {10.3847/1538-4365/ac6c03},
archivePrefix = {arXiv},
       eprint = {2203.07357},
 primaryClass = {astro-ph.GA},
       adsurl = {https://ui.adsabs.harvard.edu/abs/2022ApJS..261...31B}
}

@ARTICLE{chisholm2017,
       author = {{Chisholm}, John and {Tremonti}, Christy A. and {Leitherer}, Claus and {Chen}, Yanmei},
        title = "{The mass and momentum outflow rates of photoionized galactic outflows}",
      journal = {\mnras},
         year = 2017,
        month = aug,
       volume = {469},
       number = {4},
        pages = {4831-4849},
          doi = {10.1093/mnras/stx1164},
archivePrefix = {arXiv},
       eprint = {1702.07351},
 primaryClass = {astro-ph.GA},
       adsurl = {https://ui.adsabs.harvard.edu/abs/2017MNRAS.469.4831C}
}

@ARTICLE{hayward2017,
       author = {{Hayward}, Christopher C. and {Hopkins}, Philip F.},
        title = "{How stellar feedback simultaneously regulates star formation and drives outflows}",
      journal = {\mnras},
         year = 2017,
        month = feb,
       volume = {465},
       number = {2},
        pages = {1682-1698},
          doi = {10.1093/mnras/stw2888},
archivePrefix = {arXiv},
       eprint = {1510.05650},
 primaryClass = {astro-ph.GA},
       adsurl = {https://ui.adsabs.harvard.edu/abs/2017MNRAS.465.1682H}
}

@ARTICLE{heckman2015,
       author = {{Heckman}, Timothy M. and {Alexandroff}, Rachel M. and {Borthakur}, Sanchayeeta and {Overzier}, Roderik and {Leitherer}, Claus},
        title = "{The Systematic Properties of the Warm Phase of Starburst-Driven Galactic Winds}",
      journal = {\apj},
         year = 2015,
        month = aug,
       volume = {809},
       number = {2},
          eid = {147},
        pages = {147},
          doi = {10.1088/0004-637X/809/2/147},
archivePrefix = {arXiv},
       eprint = {1507.05622},
 primaryClass = {astro-ph.GA},
       adsurl = {https://ui.adsabs.harvard.edu/abs/2015ApJ...809..147H}
}

@ARTICLE{hopkins2014,
       author = {{Hopkins}, Philip F. and {Kere{\v{s}}}, Du{\v{s}}an and {O{\~n}orbe}, Jos{\'e} and {Faucher-Gigu{\`e}re}, Claude-Andr{\'e} and {Quataert}, Eliot and {Murray}, Norman and {Bullock}, James S.},
        title = "{Galaxies on FIRE (Feedback In Realistic Environments): stellar feedback explains cosmologically inefficient star formation}",
      journal = {\mnras},
         year = 2014,
        month = nov,
       volume = {445},
       number = {1},
        pages = {581-603},
          doi = {10.1093/mnras/stu1738},
archivePrefix = {arXiv},
       eprint = {1311.2073},
 primaryClass = {astro-ph.CO},
       adsurl = {https://ui.adsabs.harvard.edu/abs/2014MNRAS.445..581H}
}

@ARTICLE{leitherer1999,
       author = {{Leitherer}, Claus and {Schaerer}, Daniel and {Goldader}, Jeffrey D. and {Delgado}, Rosa M. Gonz{\'a}lez and {Robert}, Carmelle and {Kune}, Denis Foo and {de Mello}, Du{\'\i}lia F. and {Devost}, Daniel and {Heckman}, Timothy M.},
        title = "{Starburst99: Synthesis Models for Galaxies with Active Star Formation}",
      journal = {\apjs},
         year = 1999,
        month = jul,
       volume = {123},
       number = {1},
        pages = {3-40},
          doi = {10.1086/313233},
archivePrefix = {arXiv},
       eprint = {astro-ph/9902334},
 primaryClass = {astro-ph},
       adsurl = {https://ui.adsabs.harvard.edu/abs/1999ApJS..123....3L}
}

@ARTICLE{martin2005,
       author = {{Martin}, Crystal L.},
        title = "{Mapping Large-Scale Gaseous Outflows in Ultraluminous Galaxies with Keck II ESI Spectra: Variations in Outflow Velocity with Galactic Mass}",
      journal = {\apj},
         year = 2005,
        month = mar,
       volume = {621},
       number = {1},
        pages = {227-245},
          doi = {10.1086/427277},
archivePrefix = {arXiv},
       eprint = {astro-ph/0410247},
 primaryClass = {astro-ph},
       adsurl = {https://ui.adsabs.harvard.edu/abs/2005ApJ...621..227M}
}

@ARTICLE{murray2005,
       author = {{Murray}, Norman and {Quataert}, Eliot and {Thompson}, Todd A.},
        title = "{On the Maximum Luminosity of Galaxies and Their Central Black Holes: Feedback from Momentum-driven Winds}",
      journal = {\apj},
         year = 2005,
        month = jan,
       volume = {618},
       number = {2},
        pages = {569-585},
          doi = {10.1086/426067},
archivePrefix = {arXiv},
       eprint = {astro-ph/0406070},
 primaryClass = {astro-ph},
       adsurl = {https://ui.adsabs.harvard.edu/abs/2005ApJ...618..569M}
}

@ARTICLE{pandya2021,
       author = {{Pandya}, Viraj and {Fielding}, Drummond B. and {Angl{\'e}s-Alc{\'a}zar}, Daniel and {Somerville}, Rachel S. and {Bryan}, Greg L. and {Hayward}, Christopher C. and {Stern}, Jonathan and {Kim}, Chang-Goo and {Quataert}, Eliot and {Forbes}, John C. and {Faucher-Gigu{\`e}re}, Claude-Andr{\'e} and {Feldmann}, Robert and {Hafen}, Zachary and {Hopkins}, Philip F. and {Kere{\v{s}}}, Du{\v{s}}an and {Murray}, Norman and {Wetzel}, Andrew},
        title = "{Characterizing mass, momentum, energy, and metal outflow rates of multiphase galactic winds in the FIRE-2 cosmological simulations}",
      journal = {\mnras},
         year = 2021,
        month = dec,
       volume = {508},
       number = {2},
        pages = {2979-3008},
          doi = {10.1093/mnras/stab2714},
archivePrefix = {arXiv},
       eprint = {2103.06891},
 primaryClass = {astro-ph.GA},
       adsurl = {https://ui.adsabs.harvard.edu/abs/2021MNRAS.508.2979P}
}

@ARTICLE{scarlata2015,
       author = {{Scarlata}, C. and {Panagia}, N.},
        title = "{A Semi-analytical Line Transfer Model to Interpret the Spectra of Galaxy Outflows}",
      journal = {\apj},
         year = 2015,
        month = mar,
       volume = {801},
       number = {1},
          eid = {43},
        pages = {43},
          doi = {10.1088/0004-637X/801/1/43},
archivePrefix = {arXiv},
       eprint = {1501.07282},
 primaryClass = {astro-ph.GA},
       adsurl = {https://ui.adsabs.harvard.edu/abs/2015ApJ...801...43S}
}

@ARTICLE{shapley2003,
       author = {{Shapley}, Alice E. and {Steidel}, Charles C. and {Pettini}, Max and {Adelberger}, Kurt L.},
        title = "{Rest-Frame Ultraviolet Spectra of z\raisebox{-0.5ex}\textasciitilde3 Lyman Break Galaxies}",
      journal = {\apj},
         year = 2003,
        month = may,
       volume = {588},
       number = {1},
        pages = {65-89},
          doi = {10.1086/373922},
archivePrefix = {arXiv},
       eprint = {astro-ph/0301230},
 primaryClass = {astro-ph},
       adsurl = {https://ui.adsabs.harvard.edu/abs/2003ApJ...588...65S}
}

@ARTICLE{tumlinson2017,
       author = {{Tumlinson}, Jason and {Peeples}, Molly S. and {Werk}, Jessica K.},
        title = "{The Circumgalactic Medium}",
      journal = {\araa},
         year = 2017,
        month = aug,
       volume = {55},
       number = {1},
        pages = {389-432},
          doi = {10.1146/annurev-astro-091916-055240},
archivePrefix = {arXiv},
       eprint = {1709.09180},
 primaryClass = {astro-ph.GA},
       adsurl = {https://ui.adsabs.harvard.edu/abs/2017ARA&A..55..389T}
}

@ARTICLE{prochaska2011,
       author = {{Prochaska}, J. Xavier and {Kasen}, Daniel and {Rubin}, Kate},
        title = "{Simple Models of Metal-line Absorption and Emission from Cool Gas Outflows}",
      journal = {\apj},
         year = 2011,
        month = jun,
       volume = {734},
       number = {1},
          eid = {24},
        pages = {24},
          doi = {10.1088/0004-637X/734/1/24},
archivePrefix = {arXiv},
       eprint = {1102.3444},
 primaryClass = {astro-ph.GA},
       adsurl = {https://ui.adsabs.harvard.edu/abs/2011ApJ...734...24P}
}

@ARTICLE{xu2022,
       author = {{Xu}, Xinfeng and {Heckman}, Timothy and {Henry}, Alaina and {Berg}, Danielle A. and {Chisholm}, John and {James}, Bethan L. and {Martin}, Crystal L. and {Stark}, Daniel P. and {Aloisi}, Alessandra and {Amor{\'\i}n}, Ricardo O. and {Arellano-C{\'o}rdova}, Karla Z. and {Bordoloi}, Rongmon and {Charlot}, St{\'e}phane and {Chen}, Zuyi and {Hayes}, Matthew and {Mingozzi}, Matilde and {Sugahara}, Yuma and {Kewley}, Lisa J. and {Ouchi}, Masami and {Scarlata}, Claudia and {Steidel}, Charles C.},
        title = "{CLASSY III. The Properties of Starburst-driven Warm Ionized Outflows}",
      journal = {\apj},
         year = 2022,
        month = jul,
       volume = {933},
       number = {2},
          eid = {222},
        pages = {222},
          doi = {10.3847/1538-4357/ac6d56},
archivePrefix = {arXiv},
       eprint = {2204.09181},
 primaryClass = {astro-ph.GA},
       adsurl = {https://ui.adsabs.harvard.edu/abs/2022ApJ...933..222X}
}

@ARTICLE{hayes2023,
       author = {{Hayes}, Matthew J.},
        title = "{Accelerating galaxy winds during the big bang of starbursts}",
      journal = {\mnras},
         year = 2023,
        month = feb,
       volume = {519},
       number = {1},
        pages = {L26-L31},
          doi = {10.1093/mnrasl/slac135},
archivePrefix = {arXiv},
       eprint = {2210.11495},
 primaryClass = {astro-ph.GA},
       adsurl = {https://ui.adsabs.harvard.edu/abs/2023MNRAS.519L..26H}
}

@ARTICLE{muratov2015,
       author = {{Muratov}, Alexander L. and {Kere{\v{s}}}, Du{\v{s}}an and {Faucher-Gigu{\`e}re}, Claude-Andr{\'e} and {Hopkins}, Philip F. and {Quataert}, Eliot and {Murray}, Norman},
        title = "{Gusty, gaseous flows of FIRE: galactic winds in cosmological simulations with explicit stellar feedback}",
      journal = {\mnras},
         year = 2015,
        month = dec,
       volume = {454},
       number = {3},
        pages = {2691-2713},
          doi = {10.1093/mnras/stv2126},
archivePrefix = {arXiv},
       eprint = {1501.03155},
 primaryClass = {astro-ph.GA},
       adsurl = {https://ui.adsabs.harvard.edu/abs/2015MNRAS.454.2691M}
}

@ARTICLE{green2012,
       author = {{Green}, James C. and {Froning}, Cynthia S. and {Osterman}, Steve and {Ebbets}, Dennis and {Heap}, Sara H. and {Leitherer}, Claus and {Linsky}, Jeffrey L. and {Savage}, Blair D. and {Sembach}, Kenneth and {Shull}, J. Michael and {Siegmund}, Oswald H.~W. and {Snow}, Theodore P. and {Spencer}, John and {Stern}, S. Alan and {Stocke}, John and {Welsh}, Barry and {B{\'e}land}, St{\'e}phane and {Burgh}, Eric B. and {Danforth}, Charles and {France}, Kevin and {Keeney}, Brian and {McPhate}, Jason and {Penton}, Steven V. and {Andrews}, John and {Brownsberger}, Kenneth and {Morse}, Jon and {Wilkinson}, Erik},
        title = "{The Cosmic Origins Spectrograph}",
      journal = {\apj},
         year = 2012,
        month = jan,
       volume = {744},
       number = {1},
          eid = {60},
        pages = {60},
          doi = {10.1088/0004-637X/744/1/6010.1086/141956},
archivePrefix = {arXiv},
       eprint = {1110.0462},
 primaryClass = {astro-ph.IM},
       adsurl = {https://ui.adsabs.harvard.edu/abs/2012ApJ...744...60G}
}

@ARTICLE{james2022,
       author = {{James}, Bethan L. and {Berg}, Danielle A. and {King}, Teagan and {Sahnow}, David J. and {Mingozzi}, Matilde and {Chisholm}, John and {Heckman}, Timothy and {Martin}, Crystal L. and {Stark}, Dan P. and {Aloisi}, Alessandra and {Amor{\'\i}n}, Ricardo O. and {Arellano-C{\'o}rdova}, Karla Z. and {Bayliss}, Matthew and {Bordoloi}, Rongmon and {Brinchmann}, Jarle and {Charlot}, St{\'e}phane and {Chen}, Zuyi and {Chevallard}, Jacopo and {Clark}, Ilyse and {Erb}, Dawn K. and {Feltre}, Anna and {Hayes}, Matthew and {Henry}, Alaina and {Hernandez}, Svea and {Jaskot}, Anne and {Kewley}, Lisa J. and {Kumari}, Nimisha and {Leitherer}, Claus and {Llerena}, Mario and {Maseda}, Michael and {Nanayakkara}, Themiya and {Ouchi}, Masami and {Plat}, Adele and {Pogge}, Richard W. and {Ravindranath}, Swara and {Rigby}, Jane R. and {Scarlata}, Claudia and {Senchyna}, Peter and {Skillman}, Evan D. and {Steidel}, Charles C. and {Strom}, Allison L. and {Sugahara}, Yuma and {Wilkins}, Stephen M. and {Wofford}, Aida and {Xu}, Xinfeng and {Classy Team}},
        title = "{CLASSY. II. A Technical Overview of the COS Legacy Archive Spectroscopic Survey}",
      journal = {\apjs},
         year = 2022,
        month = oct,
       volume = {262},
       number = {2},
          eid = {37},
        pages = {37},
          doi = {10.3847/1538-4365/ac8008},
archivePrefix = {arXiv},
       eprint = {2206.01224},
 primaryClass = {astro-ph.GA},
       adsurl = {https://ui.adsabs.harvard.edu/abs/2022ApJS..262...37J}
}

@BOOK{sobolev,
       author = {{Sobolev}, V.~V.},
        title = "{Moving Envelopes of Stars}",
         year = 1960,
          doi = {10.4159/harvard.9780674864658},
       adsurl = {https://ui.adsabs.harvard.edu/abs/1960mes..book.....S}
}

@ARTICLE{carr2025,
       author = {{Carr}, Cody A. and {Cen}, Renyue and {Scarlata}, Claudia and {Xu}, Xinfeng and {Henry}, Alaina and {Marques-Chaves}, Rui and {Schaerer}, Daniel and {Amor{\'\i}n}, Ricardo O. and {Oey}, M.~S. and {Komarova}, Lena and {Flury}, Sophia and {Jaskot}, Anne and {Saldana-Lopez}, Alberto and {Ji}, Zhiyuan and {Huberty}, Mason and {Heckman}, Timothy and {{\"O}stlin}, G{\"o}ran and {Bait}, Omkar and {Hayes}, Matthew James and {Thuan}, Trinh and {Ravindranath}, Swara and {Berg}, Danielle A. and {Giavalisco}, Mauro and {Rutkowski}, Michael and {Borthakur}, Sanchayeeta and {Chisholm}, John and {Ferguson}, Harry C. and {Michel-Dansac}, Leo and {Verhamme}, Anne and {Worseck}, G{\'a}bor},
        title = "{The Effect of Radiation and Supernovae Feedback on LyC Escape in Local Star-forming Galaxies}",
      journal = {\apj},
         year = 2025,
        month = apr,
       volume = {982},
       number = {2},
          eid = {137},
        pages = {137},
          doi = {10.3847/1538-4357/adb72f},
archivePrefix = {arXiv},
       eprint = {2409.05180},
 primaryClass = {astro-ph.GA},
       adsurl = {https://ui.adsabs.harvard.edu/abs/2025ApJ...982..137C}
}

@ARTICLE{gazagnes2024,
       author = {{Gazagnes}, Simon and {Cullen}, Fergus and {Mauerhofer}, Valentin and {Begley}, Ryan and {Berg}, Danielle and {Blaizot}, Jeremy and {Chisholm}, John and {Garel}, Thibault and {Leclercq}, Floriane and {McLure}, Ross J. and {Verhamme}, Anne},
        title = "{Comparing the VANDELS Sample to a Zoom-in Radiative Hydrodynamical Simulation: Using the Si II and C II Line Spectra as Tracers of Galaxy Evolution and Lyman Continuum Leakage}",
      journal = {\apj},
         year = 2024,
        month = jul,
       volume = {969},
       number = {1},
          eid = {50},
        pages = {50},
          doi = {10.3847/1538-4357/ad47a4},
archivePrefix = {arXiv},
       eprint = {2405.03759},
 primaryClass = {astro-ph.GA},
       adsurl = {https://ui.adsabs.harvard.edu/abs/2024ApJ...969...50G}
}

@ARTICLE{xu2025,
       author = {{Xu}, Xinfeng and {Henry}, Alaina and {Heckman}, Timothy and {Carr}, Cody and {Strom}, Allison L. and {Jones}, Tucker and {Berg}, Danielle A. and {Chisholm}, John and {Erb}, Dawn and {James}, Bethan L. and {Jaskot}, Anne and {Martin}, Crystal L. and {Mingozzi}, Matilde and {Senchyna}, Peter and {Roy}, Namrata and {Scarlata}, Claudia and {Stark}, Daniel P.},
        title = "{Shining a Light on the Connections between Galactic Outflows Seen in Absorption and Emission Lines}",
      journal = {\apj},
         year = 2025,
        month = may,
       volume = {984},
       number = {1},
          eid = {94},
        pages = {94},
          doi = {10.3847/1538-4357/adc302},
archivePrefix = {arXiv},
       eprint = {2409.19776},
 primaryClass = {astro-ph.GA},
       adsurl = {https://ui.adsabs.harvard.edu/abs/2025ApJ...984...94X}
}

@ARTICLE{mcquinn2019,
       author = {{McQuinn}, Kristen. B.~W. and {van Zee}, Liese and {Skillman}, Evan D.},
        title = "{Galactic Winds in Low-mass Galaxies}",
      journal = {\apj},
         year = 2019,
        month = nov,
       volume = {886},
       number = {1},
          eid = {74},
        pages = {74},
          doi = {10.3847/1538-4357/ab4c37},
archivePrefix = {arXiv},
       eprint = {1910.04167},
 primaryClass = {astro-ph.GA},
       adsurl = {https://ui.adsabs.harvard.edu/abs/2019ApJ...886...74M}
}

@ARTICLE{jennings2025,
       author = {{Jennings}, R. Michael and {Henry}, Alaina and {Mauerhofer}, Valentin and {Heckman}, Timothy and {Scarlata}, Claudia and {Carr}, Cody and {Xu}, Xinfeng and {Huberty}, Mason and {Gazagnes}, Simon and {Jaskot}, Anne E. and {Blaizot}, Jeremy and {Verhamme}, Anne and {Flury}, Sophia R. and {Saldana-Lopez}, Alberto and {Hayes}, Matthew J. and {Trebitsch}, Maxime},
        title = "{A Simulated Galaxy Laboratory: Exploring the Observational Effects on UV Spectral Absorption Line Measurements}",
      journal = {\apj},
         year = 2025,
        month = jan,
       volume = {979},
       number = {1},
          eid = {64},
        pages = {64},
          doi = {10.3847/1538-4357/ad9b13},
archivePrefix = {arXiv},
       eprint = {2412.02794},
 primaryClass = {astro-ph.GA},
       adsurl = {https://ui.adsabs.harvard.edu/abs/2025ApJ...979...64J}
}

@ARTICLE{diamond2012,
       author = {{Diamond-Stanic}, Aleksandar M. and {Moustakas}, John and {Tremonti}, Christy A. and {Coil}, Alison L. and {Hickox}, Ryan C. and {Robaina}, Aday R. and {Rudnick}, Gregory H. and {Sell}, Paul H.},
        title = "{High-velocity Outflows without AGN Feedback: Eddington-limited Star Formation in Compact Massive Galaxies}",
      journal = {\apjl},
         year = 2012,
        month = aug,
       volume = {755},
       number = {2},
          eid = {L26},
        pages = {L26},
          doi = {10.1088/2041-8205/755/2/L26},
archivePrefix = {arXiv},
       eprint = {1205.2368},
 primaryClass = {astro-ph.CO},
       adsurl = {https://ui.adsabs.harvard.edu/abs/2012ApJ...755L..26D}
}

@ARTICLE{flury2025,
       author = {{Flury}, Sophia R. and {Jaskot}, Anne E. and {Saldana-Lopez}, Alberto and {Oey}, M.~S. and {Chisholm}, John and {Amor{\'\i}n}, Ricardo and {Bait}, Omkar and {Borthakur}, Sanchayeeta and {Carr}, Cody and {Ferguson}, Henry C. and {Giavalisco}, Mauro and {Hayes}, Matthew and {Heckman}, Timothy and {Henry}, Alaina and {Ji}, Zhiyuan and {Komarova}, Lena and {Leclercq}, Florian and {Le Reste}, Alexandra and {McCandliss}, Stephan and {Marques-Chaves}, Rui and {{\"O}stlin}, G{\"o}ran and {Pentericci}, Laura and {Ravindranath}, Swara and {Rutkowski}, Michael and {Scarlata}, Claudia and {Schaerer}, Daniel and {Thuan}, Trinh and {Trebitsch}, Maxime and {Vanzella}, Eros and {Verhamme}, Anne and {Wang}, Bingjie and {Worseck}, G{\'a}bor and {Xu}, Xinfeng},
        title = "{The Low-redshift Lyman Continuum Survey: The Roles of Stellar Feedback and Interstellar Medium Geometry in LyC Escape}",
      journal = {\apj},
         year = 2025,
        month = may,
       volume = {985},
       number = {1},
          eid = {128},
        pages = {128},
          doi = {10.3847/1538-4357/adc305},
archivePrefix = {arXiv},
       eprint = {2409.12118},
 primaryClass = {astro-ph.GA},
       adsurl = {https://ui.adsabs.harvard.edu/abs/2025ApJ...985..128F}
}

@ARTICLE{flury2022,
       author = {{Flury}, Sophia R. and {Jaskot}, Anne E. and {Ferguson}, Harry C. and {Worseck}, G{\'a}bor and {Makan}, Kirill and {Chisholm}, John and {Saldana-Lopez}, Alberto and {Schaerer}, Daniel and {McCandliss}, Stephan and {Wang}, Bingjie and {Ford}, N.~M. and {Heckman}, Timothy and {Ji}, Zhiyuan and {Giavalisco}, Mauro and {Amorin}, Ricardo and {Atek}, Hakim and {Blaizot}, Jeremy and {Borthakur}, Sanchayeeta and {Carr}, Cody and {Castellano}, Marco and {Cristiani}, Stefano and {De Barros}, Stephane and {Dickinson}, Mark and {Finkelstein}, Steven L. and {Fleming}, Brian and {Fontanot}, Fabio and {Garel}, Thibault and {Grazian}, Andrea and {Hayes}, Matthew and {Henry}, Alaina and {Mauerhofer}, Valentin and {Micheva}, Genoveva and {Oey}, M.~S. and {Ostlin}, Goran and {Papovich}, Casey and {Pentericci}, Laura and {Ravindranath}, Swara and {Rosdahl}, Joakim and {Rutkowski}, Michael and {Santini}, Paola and {Scarlata}, Claudia and {Teplitz}, Harry and {Thuan}, Trinh and {Trebitsch}, Maxime and {Vanzella}, Eros and {Verhamme}, Anne and {Xu}, Xinfeng},
        title = "{The Low-redshift Lyman Continuum Survey. I. New, Diverse Local Lyman Continuum Emitters}",
      journal = {\apjs},
         year = 2022,
        month = may,
       volume = {260},
       number = {1},
          eid = {1},
        pages = {1},
          doi = {10.3847/1538-4365/ac5331},
archivePrefix = {arXiv},
       eprint = {2201.11716},
 primaryClass = {astro-ph.GA},
       adsurl = {https://ui.adsabs.harvard.edu/abs/2022ApJS..260....1F}
}

@ARTICLE{zhu2015,
       author = {{Zhu}, Guangtun Ben and {Comparat}, Johan and {Kneib}, Jean-Paul and {Delubac}, Timoth{\'e}e and {Raichoor}, Anand and {Dawson}, Kyle S. and {Newman}, Jeffrey and {Y{\`e}che}, Christophe and {Zhou}, Xu and {Schneider}, Donald P.},
        title = "{Near-ultraviolet Spectroscopy of Star-forming Galaxies from eBOSS: Signatures of Ubiquitous Galactic-scale Outflows}",
      journal = {\apj},
         year = 2015,
        month = dec,
       volume = {815},
       number = {1},
          eid = {48},
        pages = {48},
          doi = {10.1088/0004-637X/815/1/48},
archivePrefix = {arXiv},
       eprint = {1507.07979},
 primaryClass = {astro-ph.GA},
       adsurl = {https://ui.adsabs.harvard.edu/abs/2015ApJ...815...48Z}
}

@ARTICLE{fluetsch2021,
       author = {{Fluetsch}, A. and {Maiolino}, R. and {Carniani}, S. and {Arribas}, S. and {Belfiore}, F. and {Bellocchi}, E. and {Cazzoli}, S. and {Cicone}, C. and {Cresci}, G. and {Fabian}, A.~C. and {Gallagher}, R. and {Ishibashi}, W. and {Mannucci}, F. and {Marconi}, A. and {Perna}, M. and {Sturm}, E. and {Venturi}, G.},
        title = "{Properties of the multiphase outflows in local (ultra)luminous infrared galaxies}",
      journal = {\mnras},
         year = 2021,
        month = aug,
       volume = {505},
       number = {4},
        pages = {5753-5783},
          doi = {10.1093/mnras/stab1666},
archivePrefix = {arXiv},
       eprint = {2006.13232},
 primaryClass = {astro-ph.GA},
       adsurl = {https://ui.adsabs.harvard.edu/abs/2021MNRAS.505.5753F}
}

@ARTICLE{smith2015,
       author = {{Smith}, Aaron and {Safranek-Shrader}, Chalence and {Bromm}, Volker and {Milosavljevi{\'c}}, Milo{\v{s}}},
        title = "{The Lyman {\ensuremath{\alpha}} signature of the first galaxies}",
      journal = {\mnras},
         year = 2015,
        month = jun,
       volume = {449},
       number = {4},
        pages = {4336-4362},
          doi = {10.1093/mnras/stv565},
archivePrefix = {arXiv},
       eprint = {1409.4480},
 primaryClass = {astro-ph.CO},
       adsurl = {https://ui.adsabs.harvard.edu/abs/2015MNRAS.449.4336S}
}

@ARTICLE{spilker2020,
       author = {{Spilker}, Justin S. and {Aravena}, Manuel and {Phadke}, Kedar A. and {B{\'e}thermin}, Matthieu and {Chapman}, Scott C. and {Dong}, Chenxing and {Gonzalez}, Anthony H. and {Hayward}, Christopher C. and {Hezaveh}, Yashar D. and {Litke}, Katrina C. and {Malkan}, Matthew A. and {Marrone}, Daniel P. and {Narayanan}, Desika and {Reuter}, Cassie and {Vieira}, Joaquin D. and {Wei{\ss}}, Axel},
        title = "{Ubiquitous Molecular Outflows in z > 4 Massive, Dusty Galaxies. II. Momentum-driven Winds Powered by Star Formation in the Early Universe}",
      journal = {\apj},
         year = 2020,
        month = dec,
       volume = {905},
       number = {2},
          eid = {86},
        pages = {86},
          doi = {10.3847/1538-4357/abc4e6},
archivePrefix = {arXiv},
       eprint = {2010.12591},
 primaryClass = {astro-ph.GA},
       adsurl = {https://ui.adsabs.harvard.edu/abs/2020ApJ...905...86S}
}

@ARTICLE{huberty2024,
       author = {{Huberty}, Mason and {Carr}, Cody and {Scarlata}, Claudia and {Heckman}, Timothy and {Henry}, Alaina and {Xu}, Xinfeng and {Arellano-C{\'o}rdova}, Karla Z. and {Berg}, Danielle A. and {Charlot}, St{\'e}phane and {Chisholm}, John and {Gazagnes}, Simon and {Hayes}, Matthew and {Hu}, Weida and {James}, Bethan L. and {Jennings}, R. Michael and {Leitherer}, Claus and {Martin}, Crystal L. and {Mingozzi}, Matilde and {Skillman}, Evan D. and {Sugahara}, Yuma},
        title = "{CLASSY. X. Highlighting Differences between Partial Covering and Semianalytic Modeling in the Estimation of Galactic Outflow Properties}",
      journal = {\apj},
         year = 2024,
        month = nov,
       volume = {975},
       number = {1},
          eid = {58},
        pages = {58},
          doi = {10.3847/1538-4357/ad725f},
archivePrefix = {arXiv},
       eprint = {2406.03646},
 primaryClass = {astro-ph.GA},
       adsurl = {https://ui.adsabs.harvard.edu/abs/2024ApJ...975...58H}
}

@ARTICLE{rupke2018,
       author = {{Rupke}, David S.~N.},
        title = "{A Review of Recent Observations of Galactic Winds Driven by Star Formation}",
      journal = {Galaxies},
         year = 2018,
        month = dec,
       volume = {6},
       number = {4},
          eid = {138},
        pages = {138},
          doi = {10.3390/galaxies6040138},
archivePrefix = {arXiv},
       eprint = {1812.05184},
 primaryClass = {astro-ph.GA},
       adsurl = {https://ui.adsabs.harvard.edu/abs/2018Galax...6..138R}
}

@ARTICLE{costa2014,
       author = {{Costa}, Tiago and {Sijacki}, Debora and {Haehnelt}, Martin G.},
        title = "{Feedback from active galactic nuclei: energy- versus momentum-driving}",
      journal = {\mnras},
         year = 2014,
        month = nov,
       volume = {444},
       number = {3},
        pages = {2355-2376},
          doi = {10.1093/mnras/stu1632},
archivePrefix = {arXiv},
       eprint = {1406.2691},
 primaryClass = {astro-ph.GA},
       adsurl = {https://ui.adsabs.harvard.edu/abs/2014MNRAS.444.2355C}
}

@BOOK{lamers1999,
       author = {{Lamers}, Henny J.~G.~L.~M. and {Cassinelli}, Joseph P.},
        title = "{Introduction to Stellar Winds}",
         year = 1999,
       adsurl = {https://ui.adsabs.harvard.edu/abs/1999isw..book.....L}
}

@ARTICLE{mitchell2020,
       author = {{Mitchell}, Peter D. and {Schaye}, Joop and {Bower}, Richard G. and {Crain}, Robert A.},
        title = "{Galactic outflow rates in the EAGLE simulations}",
      journal = {\mnras},
         year = 2020,
        month = may,
       volume = {494},
       number = {3},
        pages = {3971-3997},
          doi = {10.1093/mnras/staa938},
archivePrefix = {arXiv},
       eprint = {1910.09566},
 primaryClass = {astro-ph.GA},
       adsurl = {https://ui.adsabs.harvard.edu/abs/2020MNRAS.494.3971M}
}

@ARTICLE{werk2014,
       author = {{Werk}, Jessica K. and {Prochaska}, J. Xavier and {Tumlinson}, Jason and {Peeples}, Molly S. and {Tripp}, Todd M. and {Fox}, Andrew J. and {Lehner}, Nicolas and {Thom}, Christopher and {O'Meara}, John M. and {Ford}, Amanda Brady and {Bordoloi}, Rongmon and {Katz}, Neal and {Tejos}, Nicolas and {Oppenheimer}, Benjamin D. and {Dav{\'e}}, Romeel and {Weinberg}, David H.},
        title = "{The COS-Halos Survey: Physical Conditions and Baryonic Mass in the Low-redshift Circumgalactic Medium}",
      journal = {\apj},
         year = 2014,
        month = sep,
       volume = {792},
       number = {1},
          eid = {8},
        pages = {8},
          doi = {10.1088/0004-637X/792/1/8},
archivePrefix = {arXiv},
       eprint = {1403.0947},
 primaryClass = {astro-ph.CO},
       adsurl = {https://ui.adsabs.harvard.edu/abs/2014ApJ...792....8W}
}

@ARTICLE{nelson2019,
       author = {{Nelson}, Dylan and {Pillepich}, Annalisa and {Springel}, Volker and {Pakmor}, R{\"u}diger and {Weinberger}, Rainer and {Genel}, Shy and {Torrey}, Paul and {Vogelsberger}, Mark and {Marinacci}, Federico and {Hernquist}, Lars},
        title = "{First results from the TNG50 simulation: galactic outflows driven by supernovae and black hole feedback}",
      journal = {\mnras},
         year = 2019,
        month = dec,
       volume = {490},
       number = {3},
        pages = {3234-3261},
          doi = {10.1093/mnras/stz2306},
archivePrefix = {arXiv},
       eprint = {1902.05554},
 primaryClass = {astro-ph.GA},
       adsurl = {https://ui.adsabs.harvard.edu/abs/2019MNRAS.490.3234N}
}

@ARTICLE{laha2018,
       author = {{Laha}, Sibasish and {Guainazzi}, Matteo and {Piconcelli}, Enrico and {Gandhi}, Poshak and {Ricci}, Claudio and {Ghosh}, Ritesh and {Markowitz}, Alex G. and {Bagchi}, Joydeep},
        title = "{A Study of X-Ray Emission of Galaxies Hosting Molecular Outflows (MOX Sample)}",
      journal = {\apj},
         year = 2018,
        month = nov,
       volume = {868},
       number = {1},
          eid = {10},
        pages = {10},
          doi = {10.3847/1538-4357/aae390},
archivePrefix = {arXiv},
       eprint = {1809.07906},
 primaryClass = {astro-ph.HE},
       adsurl = {https://ui.adsabs.harvard.edu/abs/2018ApJ...868...10L}
}

@ARTICLE{werk2013,
       author = {{Werk}, Jessica K. and {Prochaska}, J. Xavier and {Thom}, Christopher and {Tumlinson}, Jason and {Tripp}, Todd M. and {O'Meara}, John M. and {Peeples}, Molly S.},
        title = "{The COS-Halos Survey: An Empirical Description of Metal-line Absorption in the Low-redshift Circumgalactic Medium}",
      journal = {\apjs},
         year = 2013,
        month = feb,
       volume = {204},
       number = {2},
          eid = {17},
        pages = {17},
          doi = {10.1088/0067-0049/204/2/17},
archivePrefix = {arXiv},
       eprint = {1212.0558},
 primaryClass = {astro-ph.CO},
       adsurl = {https://ui.adsabs.harvard.edu/abs/2013ApJS..204...17W}
}

@ARTICLE{falstad2015,
       author = {{Falstad}, N. and {Gonz{\'a}lez-Alfonso}, E. and {Aalto}, S. and {van der Werf}, P.~P. and {Fischer}, J. and {Veilleux}, S. and {Mel{\'e}ndez}, M. and {Farrah}, D. and {Smith}, H.~A.},
        title = "{Herschel spectroscopic observations of the compact obscured nucleus in Zw 049.057}",
      journal = {\aap},
         year = 2015,
        month = aug,
       volume = {580},
          eid = {A52},
        pages = {A52},
          doi = {10.1051/0004-6361/201526114},
archivePrefix = {arXiv},
       eprint = {1505.06934},
 primaryClass = {astro-ph.GA},
       adsurl = {https://ui.adsabs.harvard.edu/abs/2015A&A...580A..52F}
}

@ARTICLE{gonzalez2012,
       author = {{Gonz{\'a}lez-Alfonso}, E. and {Fischer}, J. and {Graci{\'a}-Carpio}, J. and {Sturm}, E. and {Hailey-Dunsheath}, S. and {Lutz}, D. and {Poglitsch}, A. and {Contursi}, A. and {Feuchtgruber}, H. and {Veilleux}, S. and {Spoon}, H.~W.~W. and {Verma}, A. and {Christopher}, N. and {Davies}, R. and {Sternberg}, A. and {Genzel}, R. and {Tacconi}, L.},
        title = "{Herschel/PACS spectroscopy of NGC 4418 and Arp 220: H$_{2}$O, H$_{2}$$^{18}$O, OH, $^{18}$OH, O I, HCN, and NH$_{3}$}",
      journal = {\aap},
         year = 2012,
        month = may,
       volume = {541},
          eid = {A4},
        pages = {A4},
          doi = {10.1051/0004-6361/201118029},
archivePrefix = {arXiv},
       eprint = {1109.1118},
 primaryClass = {astro-ph.CO},
       adsurl = {https://ui.adsabs.harvard.edu/abs/2012A&A...541A...4G}
}

@ARTICLE{herrera2020,
       author = {{Herrera-Camus}, R. and {Sturm}, E. and {Graci{\'a}-Carpio}, J. and {Veilleux}, S. and {Shimizu}, T. and {Lutz}, D. and {Stone}, M. and {Gonz{\'a}lez-Alfonso}, E. and {Davies}, R. and {Fischer}, J. and {Genzel}, R. and {Maiolino}, R. and {Sternberg}, A. and {Tacconi}, L. and {Verma}, A.},
        title = "{Molecular gas inflows and outflows in ultraluminous infrared galaxies at z {\ensuremath{\sim}} 0.2 and one QSO at z = 6.1}",
      journal = {\aap},
         year = 2020,
        month = jan,
       volume = {633},
          eid = {L4},
        pages = {L4},
          doi = {10.1051/0004-6361/201937109},
archivePrefix = {arXiv},
       eprint = {1912.05548},
 primaryClass = {astro-ph.GA},
       adsurl = {https://ui.adsabs.harvard.edu/abs/2020A&A...633L...4H}
}

@ARTICLE{stocke2013,
       author = {{Stocke}, John T. and {Keeney}, Brian A. and {Danforth}, Charles W. and {Shull}, J. Michael and {Froning}, Cynthia S. and {Green}, James C. and {Penton}, Steven V. and {Savage}, Blair D.},
        title = "{Characterizing the Circumgalactic Medium of Nearby Galaxies with HST/COS and HST/STIS Absorption-line Spectroscopy}",
      journal = {\apj},
         year = 2013,
        month = feb,
       volume = {763},
       number = {2},
          eid = {148},
        pages = {148},
          doi = {10.1088/0004-637X/763/2/148},
archivePrefix = {arXiv},
       eprint = {1212.5658},
 primaryClass = {astro-ph.CO},
       adsurl = {https://ui.adsabs.harvard.edu/abs/2013ApJ...763..148S}
}

@ARTICLE{strickland1997,
       author = {{Strickland}, D.~K. and {Ponman}, T.~J. and {Stevens}, I.~R.},
        title = "{ROSAT observations of the galactic wind in M 82.}",
      journal = {\aap},
         year = 1997,
        month = apr,
       volume = {320},
        pages = {378-394},
          doi = {10.48550/arXiv.astro-ph/9608064},
archivePrefix = {arXiv},
       eprint = {astro-ph/9608064},
 primaryClass = {astro-ph},
       adsurl = {https://ui.adsabs.harvard.edu/abs/1997A&A...320..378S}
}

@ARTICLE{peebles2014,
       author = {{Peeples}, Molly S. and {Werk}, Jessica K. and {Tumlinson}, Jason and {Oppenheimer}, Benjamin D. and {Prochaska}, J. Xavier and {Katz}, Neal and {Weinberg}, David H.},
        title = "{A Budget and Accounting of Metals at z \raisebox{-0.5ex}\textasciitilde 0: Results from the COS-Halos Survey}",
      journal = {\apj},
         year = 2014,
        month = may,
       volume = {786},
       number = {1},
          eid = {54},
        pages = {54},
          doi = {10.1088/0004-637X/786/1/54},
archivePrefix = {arXiv},
       eprint = {1310.2253},
 primaryClass = {astro-ph.CO},
       adsurl = {https://ui.adsabs.harvard.edu/abs/2014ApJ...786...54P}
}

@ARTICLE{fauchergiguere2012,
       author = {{Faucher-Gigu{\`e}re}, Claude-Andr{\'e} and {Quataert}, Eliot},
        title = "{The physics of galactic winds driven by active galactic nuclei}",
      journal = {\mnras},
         year = 2012,
        month = sep,
       volume = {425},
       number = {1},
        pages = {605-622},
          doi = {10.1111/j.1365-2966.2012.21512.x},
archivePrefix = {arXiv},
       eprint = {1204.2547},
 primaryClass = {astro-ph.CO},
       adsurl = {https://ui.adsabs.harvard.edu/abs/2012MNRAS.425..605F}
}

@ARTICLE{zhang2018,
       author = {{Zhang}, Dong},
        title = "{A Review of the Theory of Galactic Winds Driven by Stellar Feedback}",
      journal = {Galaxies},
         year = 2018,
        month = nov,
       volume = {6},
       number = {4},
          eid = {114},
        pages = {114},
          doi = {10.3390/galaxies6040114},
archivePrefix = {arXiv},
       eprint = {1811.00558},
 primaryClass = {astro-ph.GA},
       adsurl = {https://ui.adsabs.harvard.edu/abs/2018Galax...6..114Z}
}

@ARTICLE{fierlinger2016,
       author = {{Fierlinger}, Katharina M. and {Burkert}, Andreas and {Ntormousi}, Evangelia and {Fierlinger}, Peter and {Schartmann}, Marc and {Ballone}, Alessandro and {Krause}, Martin G.~H. and {Diehl}, Roland},
        title = "{Stellar feedback efficiencies: supernovae versus stellar winds}",
      journal = {\mnras},
         year = 2016,
        month = feb,
       volume = {456},
       number = {1},
        pages = {710-730},
          doi = {10.1093/mnras/stv2699},
archivePrefix = {arXiv},
       eprint = {1511.05151},
 primaryClass = {astro-ph.GA},
       adsurl = {https://ui.adsabs.harvard.edu/abs/2016MNRAS.456..710F}
}

@ARTICLE{leithererheckman1995,
       author = {{Leitherer}, Claus and {Heckman}, Timothy M.},
        title = "{Synthetic Properties of Starburst Galaxies}",
      journal = {\apjs},
         year = 1995,
        month = jan,
       volume = {96},
        pages = {9},
          doi = {10.1086/192112},
       adsurl = {https://ui.adsabs.harvard.edu/abs/1995ApJS...96....9L}
}

@ARTICLE{saldanalopez2026,
       author = {{Saldana-Lopez}, A. and {Gkini}, A. and {Hayes}, M.~J. and {Lunnan}, R. and {Carr}, C.~A. and {Huberty}, M. and {Scarlata}, C. and {Schulze}, S. and {Sollerman}, J.},
        title = "{Witnessing the onset of stellar winds in Super-Luminous Supernova Hosts: implications for star-formation-driven outflows in low and high-redshift galaxies}",
      journal = {arXiv e-prints},
         year = 2026,
        month = apr,
          eid = {arXiv:2604.13865},
        pages = {arXiv:2604.13865},
          doi = {10.48550/arXiv.2604.13865},
archivePrefix = {arXiv},
       eprint = {2604.13865},
 primaryClass = {astro-ph.GA},
       adsurl = {https://ui.adsabs.harvard.edu/abs/2026arXiv260413865S}
}

@ARTICLE{carrnature2025,
       author = {{Carr}, Cody A. and {Cen}, Renyue and {McCandliss}, Stephan and {Ford}, Jack and {Saldana-Lopez}, Alberto and {Scarlata}, Claudia and {Huberty}, Mason and {Jaskot}, Anne and {Flury}, Sophia and {Oey}, M.~S. and {Amor{\'\i}n}, Ricardo O. and {Borthakur}, Sanchayeeta and {Hayes}, Matthew and {Heckman}, Timothy and {Ji}, Zhiyuan and {Komarova}, Lena and {Le Reste}, Alexandra and {Leclercq}, Floriane and {Marques-Chaves}, Rui and {Michel-Dansac}, Leo and {{\"O}stlin}, G{\"o}ran and {Ravindranath}, Swara and {Rutkowski}, Michael J. and {Schaerer}, Daniel and {Thuan}, Trinh and {Vanzella}, Eros and {Wang}, Bingjie and {Xu}, Xinfeng},
        title = "{Supernovae Driven Winds Impede Lyman Continuum Escape from Dwarf Galaxies in First 10 Myr}",
      journal = {arXiv e-prints},
         year = 2025,
        month = oct,
          eid = {arXiv:2510.21197},
        pages = {arXiv:2510.21197},
          doi = {10.48550/arXiv.2510.21197},
archivePrefix = {arXiv},
       eprint = {2510.21197},
 primaryClass = {astro-ph.GA},
       adsurl = {https://ui.adsabs.harvard.edu/abs/2025arXiv251021197C}
}

@ARTICLE{heckman2017,
       author = {{Heckman}, Timothy and {Borthakur}, Sanchayeeta and {Wild}, Vivienne and {Schiminovich}, David and {Bordoloi}, Rongmon},
        title = "{COS-burst: Observations of the Impact of Starburst-driven Winds on the Properties of the Circum-galactic Medium}",
      journal = {\apj},
         year = 2017,
        month = sep,
       volume = {846},
       number = {2},
          eid = {151},
        pages = {151},
          doi = {10.3847/1538-4357/aa80dc},
archivePrefix = {arXiv},
       eprint = {1707.05933},
 primaryClass = {astro-ph.GA},
       adsurl = {https://ui.adsabs.harvard.edu/abs/2017ApJ...846..151H}
}

@ARTICLE{santoro2020,
       author = {{Santoro}, F. and {Tadhunter}, C. and {Baron}, D. and {Morganti}, R. and {Holt}, J.},
        title = "{AGN-driven outflows and the AGN feedback efficiency in young radio galaxies}",
      journal = {\aap},
         year = 2020,
        month = dec,
       volume = {644},
          eid = {A54},
        pages = {A54},
          doi = {10.1051/0004-6361/202039077},
archivePrefix = {arXiv},
       eprint = {2009.11175},
 primaryClass = {astro-ph.GA},
       adsurl = {https://ui.adsabs.harvard.edu/abs/2020A&A...644A..54S}
}

@ARTICLE{xu2023,
       author = {{Xu}, Xinfeng and {Heckman}, Timothy and {Henry}, Alaina and {Berg}, Danielle A. and {Chisholm}, John and {James}, Bethan L. and {Martin}, Crystal L. and {Stark}, Daniel P. and {Hayes}, Matthew and {Arellano-C{\'o}rdova}, Karla Z. and {Carr}, Cody and {Huberty}, Mason and {Mingozzi}, Matilde and {Scarlata}, Claudia and {Sugahara}, Yuma},
        title = "{CLASSY. VI. The Density, Structure, and Size of Absorption-line Outflows in Starburst Galaxies}",
      journal = {\apj},
         year = 2023,
        month = may,
       volume = {948},
       number = {1},
          eid = {28},
        pages = {28},
          doi = {10.3847/1538-4357/acbf46},
archivePrefix = {arXiv},
       eprint = {2301.11498},
 primaryClass = {astro-ph.GA},
       adsurl = {https://ui.adsabs.harvard.edu/abs/2023ApJ...948...28X}
}

@ARTICLE{peng2025,
       author = {{Peng}, Zixuan and {Martin}, Crystal L. and {Chen}, Zirui and {Fielding}, Drummond B. and {Xu}, Xinfeng and {Heckman}, Timothy and {Ramambason}, Lise and {Li}, Yuan and {Carr}, Cody and {Hu}, Weida and {Chen}, Zuyi and {Scarlata}, Claudia and {Henry}, Alaina},
        title = "{Physical Origins of Outflowing Cold Clouds in Local Star-forming Dwarf Galaxies}",
      journal = {\apj},
         year = 2025,
        month = mar,
       volume = {981},
       number = {2},
          eid = {171},
        pages = {171},
          doi = {10.3847/1538-4357/ada606},
archivePrefix = {arXiv},
       eprint = {2412.05371},
 primaryClass = {astro-ph.GA},
       adsurl = {https://ui.adsabs.harvard.edu/abs/2025ApJ...981..171P}
}

@ARTICLE{martin2024,
       author = {{Martin}, Crystal L. and {Peng}, Zixuan and {Li}, Yuan},
        title = "{Resolving the Mechanical and Radiative Feedback in J1044+0353 with Keck Cosmic Web Imager Spectral Mapping}",
      journal = {\apj},
         year = 2024,
        month = may,
       volume = {966},
       number = {2},
          eid = {190},
        pages = {190},
          doi = {10.3847/1538-4357/ad34ac},
archivePrefix = {arXiv},
       eprint = {2403.11390},
 primaryClass = {astro-ph.GA},
       adsurl = {https://ui.adsabs.harvard.edu/abs/2024ApJ...966..190M}
}

@ARTICLE{parker2025,
       author = {{Parker}, Kaelee S. and {Berg}, Danielle A. and {Chisholm}, John and {Gazagnes}, Simon and {Flury}, Sophia R. and {Carr}, Cody and {Huberty}, Mason and {Jaskot}, Anne E. and {Hayes}, Matthew J. and {Saldana-Lopez}, Alberto and {Hernandez}, Svea and {Nanayakkara}, Themiya and {James}, Bethan L. and {Arellano-C{\'o}rdova}, Karla Z. and {Strom}, Allison L. and {Senchyna}, Peter and {Mingozzi}, Matilde and {Heckman}, Timothy and {Xu}, Xinfeng and {Henry}, Alaina and {Amor{\'\i}n}, Ricardo O. and {Mauerhofer}, Valentin and {Martin}, Crystal L. and {Erb}, Dawn K. and {Skillman}, Evan D. and {Rubin}, Kate H.~R. and {Trevino}, John and {Leitherer}, Claus},
        title = "{CLASSY. XIII. Cutting through the Clouds{\textemdash}Comparing Indirect Tracers of Ionizing Photon Escape}",
      journal = {\apj},
         year = 2026,
        month = jan,
       volume = {997},
       number = {1},
          eid = {98},
        pages = {98},
          doi = {10.3847/1538-4357/ae22d9},
       adsurl = {https://ui.adsabs.harvard.edu/abs/2026ApJ...997...98P}
}

@ARTICLE{carrcen2025,
       author = {{Carr}, Cody and {Cen}, Renyue and {Flury}, Sophia and {Oey}, Sally and {McCandliss}, Stephan and {Strom}, Allison},
        title = "{How Do Ionizing Photons Escape from Star-Forming Galaxies?}",
      journal = {arXiv e-prints},
         year = 2025,
        month = jun,
          eid = {arXiv:2506.23105},
        pages = {arXiv:2506.23105},
          doi = {10.48550/arXiv.2506.23105},
archivePrefix = {arXiv},
       eprint = {2506.23105},
 primaryClass = {astro-ph.GA},
       adsurl = {https://ui.adsabs.harvard.edu/abs/2025arXiv250623105C}
}

@ARTICLE{kimb2020,
       author = {{Kim}, Chang-Goo and {Ostriker}, Eve C. and {Fielding}, Drummond B. and {Smith}, Matthew C. and {Bryan}, Greg L. and {Somerville}, Rachel S. and {Forbes}, John C. and {Genel}, Shy and {Hernquist}, Lars},
        title = "{A Framework for Multiphase Galactic Wind Launching Using TIGRESS}",
      journal = {\apjl},
         year = 2020,
        month = nov,
       volume = {903},
       number = {2},
          eid = {L34},
        pages = {L34},
          doi = {10.3847/2041-8213/abc252},
archivePrefix = {arXiv},
       eprint = {2010.09090},
 primaryClass = {astro-ph.GA},
       adsurl = {https://ui.adsabs.harvard.edu/abs/2020ApJ...903L..34K}
}
\bibliographystyle{aasjournalv7}
%\bibliographystyle{apj} 
%\bibliography{../myreferences}{}

\end{document}